\documentclass[12pt]{article}
\usepackage{amsfonts,amsmath,amssymb,bbm}
\usepackage{cite,graphicx,color}
\usepackage[colorlinks=true, linkcolor=blue, citecolor=blue, linktoc=all]{hyperref}
\usepackage[usenames,dvipsnames]{xcolor}
\usepackage{diagbox}
\usepackage{extarrows}
\usepackage{url}
\usepackage[font=small,labelfont=bf,tableposition=top]{caption}
\usepackage{esint} 
\usepackage{psfrag}
\usepackage{subfigure}
\usepackage{tabularx}
\usepackage{footmisc}
\usepackage{float}
\newcommand\cc[1]{#1^{^{\kern-6pt \circ}}\kern2pt}

\renewcommand{\a}{\alpha}

\newcommand{\m}{\mu}
\newcommand{\n}{\nu}

\def\be{\begin{equation}}
\def\ee{\end{equation}}
\def\bea{\begin{eqnarray}}
\def\eea{\end{eqnarray}}
\def\ba{\begin{array}}
\def\ea{\end{array}}
\def\bi{\begin{itemize}}
\def\ei{\end{itemize}}

\newcommand{\beq}{\begin{equation}}
\newcommand{\eeq}{\end{equation}}
\newcommand{\beqn}{\begin{eqnarray}}
\newcommand{\eeqn}{\end{eqnarray}}
\newcommand{\bga}{\begin{align}}

\def\dalemb#1#2{
	{\vbox{
		\hrule height .#2pt
		\hbox{\vrule width.#2pt height#1pt \kern#1pt\vrule width.#2pt}
		\hrule height.#2pt}}
	}

  \makeatletter
\def\@fnsymbol#1{\ensuremath{\ifcase#1\or \ddagger\or
   \mathsection\or \mathparagraph\or \|\or **\or \dagger\dagger
   \or \ddagger\ddagger \else\@ctrerr\fi}}
    \makeatother

\DeclareCaptionLabelFormat{andtable}{#1~#2  \&  \tablename~\thetable}

\newcommand{\thistitle}{{\bf Remarks on the quasinormal modes of Taub-NUT-AdS$_4$}}

\begin{document}

\allowdisplaybreaks
\title{\thistitle}
\author{
	 Georgios Kalamakis\thanks{georkal@hotmail.gr}  \quad and \,\, Anastasios C. Petkou\thanks{petkou@physics.auth.gr}
	\\
\\
{\small \emph{Laboratory of Theoretical Physics}}\\{\small\emph{Aristotle University of Thessaloniki}} \\{\small\emph{54124 Thessaloniki, Greece}}}
	
%{\small \emph{\auth}}\\ }
%\date{\today}
\maketitle
\vspace{-5ex}
\begin{abstract}
\vspace{0.4cm}
We present results for the numerical evaluation of scalar quasinormal modes in Taub-NUT-AdS$_4$ spacetimes. To achieve this we consider angular modes that correspond to non-unitary highest weight $SU(2)$ representations since global regularity is not consistent with the presence of complex quasinormal modes. We show that for any non-zero value of the NUT charge $n$ there exists a region in the complex plane that contains only stable quasinormal modes. The radius of this region increases with the horizon distance and decreases with $n$ towards a constant value for infinite $n$. We also find analytic and numerical evidence for the existence of a critical NUT charge $n_{cr}$ beyond which the lowest lying stable quasinormal modes become overdamped. Our results draw an intricate picture for the holographic fluid of Taub-NUT-AdS$_4$.

\end{abstract}

\section{Introduction}

The duality between asymptotically Anti-de Sitter spacetimes and relativistic hydrodynamical systems (see \cite{Hubeny:2011hd} for a review and references) is one of the most remarkable spinoffs of AdS/CFT holography and has generated among others an enormous interest in calculations of quasinormal modes (QNMs) for a wealth of black hole perturbations. Classical reviews on the subject include \cite{Berti:2009kk, Konoplya:2011qq}.

Typically, the QNMs of asymptotically Anti-de Sitter black holes correspond to poles of the retarded Green's function of the dual holographic fluid system, and hence they are relevant to its dissipative properties. In this context, four-dimensional asymptotically AdS black holes are particularly interesting as they give rise to dissipative strongly coupled systems in 2+1 dimensions which might be experimentally probed \cite{Hartnoll:2016apf}. The seminal works on the QNMs of four-dimensional Schwarzchild AdS black holes \cite{Horowitz:1999jd, Cardoso:2001bb, Cardoso:2003cj, Michalogiorgakis:2006jc} were followed by many studies of rotating \cite{Cardoso:2013pza, Uchikata:2009zz} and charged \cite{Wang:2000gsa,Wang:2000dt,Berti:2003ud,Ficek:2023imn} asymptotically AdS black holes. Less explored are cases with nontrivial topology, of which the  prime example is the Taub-NUT Anti-de Sitter (TNAdS) spacetime. The first effort towards the calculation of QNMs in TNAdS$_4$ was done in \cite{Kalamakis:2020aaj} and it was followed by the works \cite{Cano:2021qzp,Al-Badawi:2021wdm}. 

The main complication with the holographic interpretation of TNAdS$_4$ spacetimes arises due to the presence of the Misner string\footnote{Recent works that propose a "rehabilitation" of TNAdS spacetimes include \cite{Clement:2015cxa,Kubiznak:2019yiu}.}  \cite{Misner:1963fr,Bonnor:1969ala,Astefanesei:2004kn}. At the level of perturbations around the background geometry the issue arises when one tries to solve the angular equation. For scalar perturbations this equation coincides with the equation that gives rise to the monopole harmonics \cite{Wu:1976ge} or equivalently the spin-weighted spherical harmonics \cite{Dray:1984gy,Casals:2018cgx} and its regular solutions require the quantization of QNMs in units of the NUT charge \cite{Kalamakis:2020aaj}. This is equivalent to removing the Misner string at the expense of making periodic the time coordinate. Nevertheless, it was also argued in \cite{Kalamakis:2020aaj} that one might try to relax the regularity condition for the solutions of the angular equation which would be equivalent to allowing for a physical Misner string in the bulk. Such an approach would correspond to the presence of a vortex in the boundary fluid, having anyonic properties. For a recent discussion of the angular equation in Taub NUT spacetimes see \cite{Willenborg:2024nwn}. 

In this note, following \cite{Kalamakis:2020aaj}, we elaborate on the numerical evaluation of the  QNMs of TNAdS$_4$ assuming the presence of a physical Misner string. In this case we cannot use regularity arguments to fix the eigenvalues of the angular equation, since the former condition is incompatible with the presence of complex QNMs. This should be contrasted with similar calculations of the angular modes in rotating black holes in \cite{Cardoso:2013pza, Uchikata:2009zz}. In the TNAdS$_4$ case, using the $SU(2)$ symmetry of the boundary metric as a guiding principle, we classify the angular modes into $SU(2)$ representations. These are, however, non-unitary and generically multivalued which would imply an anyonic behaviour of the holographic fluid parametrised by an arbitrary real parameter ${\cal N}$. We believe that the above properties of the angular modes are necessary for a possible holographic interpretation of  TNAdS$_4$ spacetime as a dissipative fluid. In fact, one might compare the non-unitarity property of the $SU(2)$ representations with the corresponding property of the angular modes in rotating spacetimes, such as the Kerr-AdS$_4$, where the presence of complex QNMs is tied to complex eigenvalues of the angular equation \cite{Dias:2013sdc} and appears to be closely connected with the chaotic properties of the holographic boundary fluid \cite{Blake:2021hjj,Chu:2024iie}. 

We then show that for any non-zero value of the NUT charge $n$ there exists a region in the complex plane that contains only stable quasinormal modes. The size of this {\it stable QNM region} is proportional to the horizon distance $r_+$, hence larger TNAdS$_4$ black holes are more stable. When $n$ increases the size of the stable QNM region decreases as $1/n$, but we find that it never shrinks to zero and tends to a constant value for infinite $n$. The behaviour of the quasinormal modes inside the stable QNM region depends crucially on $n$. For $n\ll 1$ the modes approach continuously the corresponding quasinormal modes of the Schwarzchild AdS$_4$ black hole. However, we observe that there exists a critical value for the NUT charge $n_{cr}$ above which the real part of the lowest quasinormal mode falls abruptly to a near zero value and approaches zero for $n\gg n_{cr}$. The imaginary part of the lowest lying QNM also decreases abruptly as $n$ approaches $n_{cr}$ from above, but remains well above zero and moreover it starts increasing for $n>n_{cr}$. A similar behaviour is observed for the higher overtones. This picture does not seem to be affected by the anyonic parameter ${\cal N}$. Remarkably, the above critical value $n_{cr}$ of the NUT charge seems to be rather insensitive to the horizon radius $r_{+}$. We are then able to present a quantitative analytic argument what this is the case. Our results show that insisting on the  presence of a physical Misner string leads to an intricate holographic fluid for TNAdS$_4$ whose salient property is the existence of a critical $n_{cr}$ which separates a propagating from an overdamped phase. The alternative holographic interpretation would be that a superfluid with quantized vortices as we suggest. 
In Section 2 we setup the calculation by briefly reviewing  the results of \cite{Kalamakis:2020aaj}. To calibrate our results we also review the well-known results for the QNMs of Schwarzchild AdS$_4$. In Section 3 we present the numerical methods of our calculations. In Section 4 we present and discuss our results.

\section{Scalar fluctuations in Schwarzchild AdS$_4$ and Taub-NUT AdS$_4$ geometries.}

The Lorentzian Schwarzschild and Taub-NUT AdS$_4$ metrics with spherical horizons can be collectively presented as\footnote{Throughout the paper we use $x^{(A,B)}=(r,t,\theta,\phi)$ and $x^{(\m,\n)}=(t,\theta,\phi)$.}
\begin{align}
\label{metric}
ds^2 = g_{AB}dx^A dx^B=\frac{dr^2}{V(r,n)} + G(r,n)\left[d\theta^2 + \sin^2\theta d\phi^2\right]-V(r,n)[dt +b(\theta,n)d\phi]^2 
\end{align}
with 
\begin{align}
V(r,n) &= \frac{1}{r^2+n^2}\left[r^2-n^2-2Mr+\frac{1}{L^2} (r^4+6n^2 r^2-3n^4)\right] \,,\\
G(r,n) &= r^2+n^2\,,\,\,\,\,\,b(n,\theta) = 2n(1-\cos\theta)\,.
\end{align}
$L$ is the AdS radius, $M>0$ is the mass parameter and $n$ is the NUT charge which is taken to be a real number. Schwarzchild SAdS$_4$ (SAdS$_4$) arises as a smooth limit of (\ref{metric}) for $n=0$.

%\item Kerr:
%\begin{align}
%F(r,\theta) &= \frac{u(r,\theta)}{w(r)} \\
%G_1(r,\theta) &= \frac{u(r,\theta)}{v(\theta} \\
%G_2(r,\theta) &= \frac{(\alpha^2+r^2)^2}{1-\frac{\alpha^2}{L^2}}\sin^2\theta \cdot \frac{v^2(\theta)}{u^2(r,\theta)} \\
%H_1(r,\theta) &= \frac{w(r)-\alpha^2\sin^2\theta v(\theta)}{u(r,\theta)} \\
%H_2(r,\theta) &= \frac{w(r)\sin^2\theta - 2\alpha (\alpha^2+r^2)v(\theta)}{(1-\frac{\alpha^2}{L^2})[w(r)-\alpha^2 v(\theta)\sin^2\theta]} \\
%H_3(r,\theta) &= \frac{\alpha^2 w(r) \sin^4\theta \sqrt{u(r,\theta)}}{(1-\frac{\alpha^2}{L^2})[w(r)-\alpha^2 v(\theta) \sin^2\theta]}
%\end{align}
%where
%\begin{align}
%u(r,\theta) &= r^2+\alpha^2\cos^2\theta \\
%v(\theta) &= 1-\frac{\alpha^2\cos^2\theta}{L^2} \\
%w(r) &= (\alpha^2+ r^2)(1+\frac{r^2}{L^2})-2Mr
%\end{align}
%\end{itemize}
For generic values of  $M>0$ and $n$ the above metric has an outer horizon at $r_+$ with the topology of $S^2$. Its position is determined by $V(r_+,n)=0$ as
\begin{align}
\label{defr+}
r_+[r_+^3+(6n^2+L^2)r_+-2ML^2]=3n^4+n^2L^2\,.
\end{align} 

AdS/CFT proposes that the vacuum spacetimes (\ref{metric}) correspond to hydrodynamic systems living in their asymptotic timelike boundaries having a conserved, symmetric and traceless energy momentum tensor whose expectation value in the fluid state is given as
 \cite{Leigh:2011au,Leigh:2012jv} 
\begin{align}
\label{bT}
&T_{\m\n}=p[3u_\m u_\n+g_{\m\n}]\,,\,\,\,p=\frac{M}{8\pi G_4 L^2}\,,\,\,\,\m,\n=0,1,2\,,\\
&\label{Tconserv}
\nabla^\m T_{\m\n}=g^{\m\n}T_{\m\n}=0\,,\,\,T_{\m\n}=T_{\n\m}\,.
\end{align}
with $G_4$ the four-dimensional Newton's constant. The velocity $u_\m$ of the boundary fluid's flow is a geodesic, shearless and expansionless congruence of the boundary metric 
\be
\label{bg}
ds^2_{bdy}=g_{\m\n}dx^\m dx^\n=-[dt+2n(1-\cos\theta)d\phi]^2+L^2d\Omega_2^2\,.
\ee
Due to the presence of the NUT parameter $n$ the flow has nonzero vorticity\footnote{Recall the definitions of acceleration $\a_\m=u^\n\nabla_\n u_\m$, expansion $\Theta=\nabla_\m u^\m$, shear $\sigma_{\m\n}=h_\m^{\,\,\sigma}h_{\n}^{\,\,\rho}(\nabla_\sigma u_\rho+\nabla_\rho u_\sigma)/2-h_{\m\n}h^{\sigma\rho}(\nabla_\sigma u_\rho)/2$ and vorticity $\omega_{\m\n}=h_\m^{\,\,\sigma}h_{\n}^{\,\,\rho}(\nabla_\sigma u_\rho-\nabla_\rho u_\sigma)/2$, where $h_{\m\n}=g_{\m\n}+u_\m u_\n$. One might interpret $u$ as a gauge field, $\omega$ as the corresponding field strength, and the circulation ${\cal C}$ (eq. \eqref{totalfluxTN}) as a charge.}
\begin{align}
\label{velocity}
&u^\m=(1,0,0)\,,\,\,u_\m=(-1,0,-2n(1-\cos\theta))\,,\\
\label{vorticity}
&u^\n\nabla_\n u_\m=\nabla_\m u^\m=\sigma_{\m\n}=0\,,\\
&\omega^{TN}_{\m\n}=\left(\begin{array}{ccc}0&0&0\\0&0&-n\sin\theta\\0&n\sin\theta&0\end{array}\right)\,.
\end{align}
Having at hand the vorticity one can calculate its flux through a bounded surface i.e. the circulation. For inviscid and barotropic fluids the circulation along any closed path is constant and hence one would expect from Stoke's theorem that the vorticity flux through a closed surface is zero. This can be verified for example in the case of the rotating boundary fluid of the Kerr-AdS$_4$ metric \cite{Kalamakis:2020aaj}. However, in the TNAdS$_4$ case the flow velocity (\ref{velocity}) is singular\footnote{This can be seen 
 as 
$\hat{u} \xrightarrow{\theta\to 0} -dt+\theta n L\hat{e}^\phi+O(\theta^3)\,,\,\,\,\hat{u} \xrightarrow{\theta\to \pi} -dt+\frac{nL}{\pi -\theta}\hat{e}^\phi+O(\pi-\theta)\,,\,\,\,\,\,\,\hat{e}^\phi=\sin\theta d\phi\,.$}
at $\theta=\pi$ and one needs to invoke Stoke's theorem for non simply connected domains. The result is that the total circulation, or equivalently the vorticity flow over the $S^2$ boundary surface,  does not vanish and it is given by 
\be
\label{totalfluxTN}
{\cal C}^{TN}_{tot}=2\oiint_{{\cal S}}dS^{\m\n}\omega^{TN}_{\m\n}=-2n\int_{0}^{2\pi}\!\!d\phi\int_0^\pi d\theta\,\sin \theta =-8\pi n\,.
\ee
Finally, since the metric (\ref{bg}) is not conformally flat, it gives rise to a non-zero Cotton tensor that has the from of a  perfect conformal fluid \cite{Mukhopadhyay:2013gja}.
\be
\label{TNCotton}
C_{\m\n}=\frac{n}{L^4}\left(1+\frac{4n^2}{L^2}\right)[3u_\m u_\n+g_{\m\n}]\,,\,\,\,\nabla^{\m}C_{\m\n}=g^{\m\n}C_{\m\n}=0\,.
\ee

\section{The quasinormal modes of TNAdS$_4$}

The holographic interpretation of TNAdS$_4$ is, to our knowledge, still an open issue.  While $T_{\m\n}$ gives the expectation value of the energy momentum tensor in a thermal state of the boundary system, the Cotton tensor $C_{\m\n}$ arises as a property of the boundary metric (\ref{bg}) which in the context of AdS/CFT is interpreted as an external source of the boundary effective action. We therefore believe that the natural interpretation of  $C_{\m\n}$ is that of a source for an operator in the boundary system, which is probably nonlocal and similar to the monopole operator in electromagnetism. Hence, the holographic boundary fluid appears to be more intricate than the the rotating fluid of e.g. Kerr-AdS$_4$. 

Nevertheless, whatever the holographic interpretation of the NUT charge $n$ is, one can easily verify that the TNAdS$_4$ metric becomes Schwarzchild AdS$_4$ in the limit $n\rightarrow 0$ and that the limit is geometrically smooth. Moreover, there does not appear to exist an upper bound on $n$ similar to the upper bound of the rotation parameter in Kerr-AdS$_4$, nor is there a useful extremal limit such as in the case of charged AdS black holes. We hope to shed partial light on these issues by studying the scalar quasinomal modes of the Taub-NUT AdS$_4$ metric. For simplicity we consider a minimally coupled scalar field $\Phi$ whose equation of motion on the fixed metric (\ref{metric}) is 
\be
\label{Phieom}
\frac{1}{\sqrt{-g}}\partial_A \left[\sqrt{-g}g^{AB}\partial_B\Phi\right]  = 0\,.
\ee
This equation is separable and assuming $\Phi(t,r,\theta,\phi) = e^{-i\omega t}R(r)Y(\theta, \phi)$ one obtains the set of equations
\begin{subequations}
\begin{align}\label{TN1}
&\left\{\frac{d}{dr}[G(r,n)V(r,n)\frac{d}{dr}] +\omega^2\frac{G(r,n)}{V(r,n)} + 4n^2\omega^2 -C\right\} R(r) = 0\,, \\
&\left\{{\bf L}^2 -C\right\}Y(\theta,\phi) = 0\,, \label{TN2}
\end{align}
\end{subequations}
where 
\begin{align}
{\bf L}^2 = - \frac{1}{\sin^2\theta}\left[\sin\theta\frac{\partial}{\partial\theta}\left(\sin\theta\frac{\partial}{\partial\theta}\right) + \left[\partial_\phi+i2n\omega(1-\cos\theta)\right]^2\right] + 4n^2\omega^2\,.
\end{align}
The separation constant $C$ corresponds to the quadratic Casimir of an $SU(2)$ representation, see \cite{Kalamakis:2020aaj} and the discussion below. 

From the outset equations (\ref{TN1}) and (\ref{TN2}) look very similar to the set of Teukolsky equations for Kerr perturbations e.g. \cite{Cardoso:2013pza}. However, contrary to these cases TNAdS$_4$ geometry possesses a richer geometrical structure which should play a role in the analysis of the fluctuation equations above. Explicitly, the boundary metric (\ref{bg}) has an $SU(2)\times \mathbb{R}$ isometry generated by the vector fields\footnote{The interested reader is asked to look in \cite{Kalamakis:2020aaj} for more details.}
\beqn
\label{TNAdS4Kill1}
{\boldsymbol \xi}_1 &=& -\sin\phi \cot\theta\,\partial_\phi +\cos\phi\,\partial_\theta -2n\sin\phi\frac{1-\cos\theta}{\sin\theta}\,\partial_t\,, \\
{\boldsymbol \xi}_2 &=& \cos\phi \cot\theta\,\partial_\phi + \sin\phi\,\partial_\theta +2n\cos\phi\frac{1-\cos\theta}{\sin\theta}\,\partial_t\,, \\
{\boldsymbol \xi}_3 &=& \partial_\phi -2n\partial_t,\qquad\qquad {\bf e} = \partial_t
\eeqn
with 
\beq
[{\boldsymbol \xi}_i,{ \boldsymbol \xi}_j ] =-\epsilon_{ijk}{ \boldsymbol\xi}_k\qquad[{\boldsymbol \xi}_i,{\bf e}]=0\quad i,j,k=1,2,3. 
\eeq
We conclude that the angular eigenfunctions $Y(\theta,\phi)$ can be classified according to representations of $SU(2)$, in contrast to the Kerr-AdS$_4$ case where there is no such requirement. Notice that when $n=0$ the angular equation (\ref{TN2}) becomes the equation that gives the usual spherical harmonics which are the unitary finite dimensional representations of $SU(2)$. To proceed we can define the complex generators as 
\be
{\bf L}_3=-i{\boldsymbol \xi}_3=-i(\partial_\phi-2 n\partial_t)\,,\,\,\,\,{\bf L}_\pm =\pm{\boldsymbol \xi}_1+ i {\boldsymbol \xi}_2 \,, \ee
which satisfy  
\begin{align}
[ {\bf L}_+, {\bf L}_-]=2  {\bf L}_3 \qquad [ {\bf L}_3, {\bf L}_\pm] = \pm  {\bf L}_\pm\,.
\end{align}
It is then straightforward to verify that the quadratic $SU(2)$ Casimir is given by 
\begin{align}
{\bf L}^2 = 
-\sum_i{\boldsymbol \xi}_i^2\,.
\end{align}
It is convenient to write the eigenvalue of the quadratic Casimir as $C=q(q+1)$ and look for solutions of the angular equation on which ${\bf L}_{\pm}$ act as raising/lowering operators. Such solutions are the eigenfunctions of the "magnetic" operator ${\bf L}_3=-i{\boldsymbol \xi}_3$. Namely we ask that 
\be
\label{L2L3eigen}
{\bf L}^2 Y(\theta,\phi)=C \,Y(\theta,\phi),\qquad
{\bf L}_3Y(\theta,\phi)=m\,Y(\theta,\phi)\,.
\ee
Due to the form of ${\bf L}_3$, to satisfy the second of (\ref{L2L3eigen}) we need to have 
\beq
\label{PhiSep}
Y(\theta,\phi)\equiv Y_{q,m,\Omega}(\theta,\phi)=Y_{q,m,\Omega}(\theta)e^{i{\cal N}\phi},
\eeq
where we have set ${\cal N}=m-\Omega$ and $\Omega=2n\omega$. This form of the angular eigenfunctions is a salient feature of the TNAdS$_4$ fluctuations if we insist that they should fall into $SU(2)$ representations. Namely, the azimuthal $\phi$-dependance of the fluctuations depends on the mode frequency $\omega$.  
Given the form of \eqref{PhiSep}  eq. \eqref{TN2}  becomes
\beq
\label{angularEq2}
\left\{\frac{1}{\sin\theta}\frac{d}{d\theta}\left[\sin\theta\frac{d}{d\theta}\right]-\frac{(m-\Omega\cos\theta)^2}{\sin^2\theta}-\Omega^2+q(q+1)\right\}Y_{q,m,\Omega}(\theta)=0\,.
\eeq
A few remarks are in order here. Eq. (\ref{angularEq2})  coincides with the equation satisfied by the {\it spin-weighted spheroidal harmonics} \cite{Berti:2005gp} if we were to set $\Omega\equiv s=0,\pm 1,\pm2,..$. This quantization condition on $\Omega$ would give us a handle on the functions $Y(\theta,\phi)$ which can then be identified with the everywhere regular spin-weighted spherical harmonics and can be constructed from the usual spherical harmonics by repeated applications of spin-raising operators \cite{Shah:2015sva}. This approach is equivalent to the quantum mechanical description of a particle with charge $e\equiv\omega$ in a magnetic monopole background, with $g\equiv-2n$ the monopole charge \cite{Balian:2005joa}.

However, the intricacy of our problem lies in the fact that $\Omega=2n\omega$ in not necessary quantized. In fact, its is independently determined by the radial equation (\ref{TN1}) which will generically give complex values for $\omega$ if we impose the usual infalling boundary condition on $r_+$. We are therefore forced to conclude that the regularity of the angular modes is incompatible with the presence of complex quasinormal modes. For that reason we developed in \cite{Kalamakis:2020aaj} an approach to solve the angular equation (\ref{angularEq2}) that allows for complex quasinormal modes $\Omega$ at the expense of the regularity of the angular eigenfunctions. Let us briefly review this here. Introducing the coordinate $u=\sin^2(\theta/2)$ (\ref{angularEq2}) becomes
\beq
\label{angularEq3}
\left\{\frac{d}{du}\left[u(1-u)\frac{d}{du}\right]+\left[q(q+1)-\Omega^2-\frac{({\cal N}+2u\Omega)^2}{4u(1- u)}\right]\right\}Y_{q,m,\Omega}(u)=0\,,
\eeq
whose solutions are hypergeometric functions of the form
\beq\label{genSolPhi}
Y_{q,m,\Omega}(u)=u^{a/2}(1-u)^{b/2}{}_2F_1(1+q+\tfrac{a+b}{2},-q+\tfrac{a+b}{2};1+a;u)\,.
\eeq
The parameters $a,b$ take  four possible values  $a=\pm{\cal N}\,,\, b=\pm(2m-{\cal N})$, however the resulting four hypergeometric functions are not all independent and it suffices to consider
\beq
Y^\pm_{q,m,\Omega}(u)=u^{\pm{\cal N}/2}(1-u)^{\pm(2m-{\cal N})/2}{}_2F_1(1+q\pm m,-q\pm m;1\pm {\cal N};u)\,.
\eeq
Overall the solutions to the scalar wave equation take the form
\beqn
\label{sepvar}
\Phi^\pm_{q,m,\Omega}(t,r,u,\phi)&=&R_{q,\Omega}(r)\Psi^\pm_{q,m,\Omega}(t,u,\phi)\,,\\
\Psi^\pm_{q,m,\Omega}(t,u,\phi)&=&e^{-i\omega t}Y^\pm_{q,m,\Omega}(u,\phi)=e^{-i\omega t}e^{i{\cal N}\phi}Y^\pm_{q,m,\Omega}(u)\,,
\eeqn
and satisfy 
\beqn
&i{\bf e}(\Psi^\pm_{q,m,\Omega}(t,u,\phi))=\omega\Psi^\pm_{q,m,\Omega}(t,u,\phi)\,,\\
&{\bf L}_3(\Psi^\pm_{q,m,\Omega}(t,u,\phi))=m\Psi^\pm_{q,m,\Omega}(t,u,\phi)\,,\\
& {\bf L}^2(\Psi^\pm_{q,m,\Omega}(t,u,\phi))=q(q+1)\Psi^\pm_{q,m,\Omega}(t,u,\phi)\,.
\eeqn
The action of the raising/lowering operators ${\bf L}_{\pm}$ on the $SU(2)$ modules can be  found using their explicit forms 
\beq
{\bf L}_\pm = \frac{ie^{\pm i\phi}}{\sqrt{u(1- u)}}\left[ \mp iu(1- u)\partial_u +\frac{1-2 u}{2}\partial_\phi+2nu\partial_t\right] \,,
\eeq
as
\beqn
\label{su2pluslower}
 {\bf L}_-(\Psi^+_{q,m,\Omega}(t,u,\phi))&=&-{\cal N}\Psi^+_{q,m-1,\Omega}(t,u,\phi)\,,\\
\label{su2up}
{\bf L}_+(\Psi^+_{q,m,\Omega}(t,u,\phi))&=&\frac{(1+q+m)(m-q)}{1+{\cal N}}\Psi^+_{q,m+1,\Omega}(t,u,\phi)\,,\\
\label{su2down}
{\bf L}_-(\Psi^-_{q,m,\Omega}(t,u,\phi))&=&\frac{(1+q-m)(m+q)}{1-{\cal N}}\Psi^-_{q,m-1,\Omega}(t,u,\phi\,,)\\
\label{su2minusraise}
 {\bf L}_+(\Psi^-_{q,m,\Omega}(t,u,\phi))&=&-{\cal N}\Psi^-_{q,m+1,\Omega}(t,u,\phi)\,.
\eeqn
%From these, we can also deduce that
%\beqn
% {\bf L}_+{\bf L}_-(\Psi^\pm_{q,m,\Omega}(t,u,\phi))&=&\Big(q(q+1)-m(m-1)\Big)\Psi^\pm_{q,m,\Omega}(t,u,\phi)\,,\\
%{\bf L}_- {\bf L}_+(\Psi^\pm_{q,m,\Omega}(t,u,\phi))&=&\Big(q(q+1)-m(m+1)\Big)\Psi^\pm_{q,m,\Omega}(t,u,\phi)\,.
%\eeqn
Eqs (\ref{su2pluslower})-(\ref{su2minusraise}) are just the known relations satisfied by $SU(2)$ representations. The difference wrt to the standard quantum mechanical case are that i) the ${\bf L}_3$ eigenvalue ${\cal N}$ is not necessarily integer, and ii) the $SU(2)$ modules are non necessarily finite dimensional. 

To proceed we make two further assumptions. Firstly we assume that ${\cal N}=m-\Omega\in{\mathbb R}$. The physical implication of our assumption is that the angular modes are {\it anyonic} and we want to attribute such a property to the presence of physical Misner string. Secondly, since the solutions of (\ref{angularEq2}) span an $SU(2)$ representation space, we can arrange them in highest/lowest weight modules. To do that notice that if $Y_{q,m,\Omega}(u,\phi)$ is a solution of eq. \eqref{TN2} corresponding to the quasinormal mode $\omega$,  then $Y^*_{q,m,\Omega}(u,\phi)\equiv \tilde Y_{\bar q,\bar m,\bar\Omega}(u,\phi)$ solves the same equation for values of the parameters $(q^*,-m^*,-\Omega^*)\equiv (\bar q,\bar m,\bar\Omega)$, and ${\bar{\cal N}}=-{\cal N}$. The two complex conjugate solutions are
\beqn
Y_{q,m,\Omega}(u,\phi)&=&
e^{i{\cal N}\phi} u^{{\cal N}/2}(1-u)^{(2m-{\cal N})/2} {}_2F_1(1+q+m,-q+m,1+m-\Omega;u)\,,
\\
\tilde Y_{\bar q,\bar m,\bar\Omega}(u,\phi)&=&e^{i\bar{\cal{N}}\phi} u^{-\bar{\cal N}/2}(1-u)^{-(2\bar m-\bar{\cal N})/2} {}_2F_1(1+\bar q-\bar m,-\bar q-\bar m,1-\bar m+\bar\Omega;u)\,.
\eeqn
We then obtain
\beq
{\bf  L}_-\left(e^{-i\bar{\omega}t}\tilde Y_{\bar q,\bar m,\bar\Omega}(u,\phi)\right)=\frac{(1+\bar q-\bar m)(-\bar q-\bar m)}{1-\bar m+\bar\Omega}e^{-i\bar{\omega}t}\tilde Y_{\bar q,\bar m-1,\bar\Omega}(u,\phi)\,,
\eeq
and
\beq
{\bf L}_+\left(e^{-i\omega t}Y_{q,m,\Omega}(u,\phi)\right)=\frac{(1+q+m)(m-q)}{1+m-\Omega}e^{-i\omega t}Y_{q,m+1,\Omega}(u,\phi)\,.
\eeq
 Therefore we can associate $Y_{q,q,\Omega}(u,\phi)$ to a highest weight state $\Psi_{q,q,\Omega}(t,u,\phi)$ that is annihilated by the raising operator as ${\bf L}_+\Psi_{q,q,\Omega}(t,u,\phi)=0$, while $\tilde Y_{\bar q,-\bar q,\bar\Omega}(u,\phi)$  to a corresponding lowest weight state  with ${\bf L}_-\tilde{\Psi}_{\bar{q},-\bar{q},\bar{\Omega}}(t,u,\phi)=0$. The highest/lowest weight modules are infinite dimensional, hence they form non-unitary representations of $SU(2)$. To every highest/lowest weight module correspond the quasinormal modes $\omega/-\omega^*$ respectively. Notice that the finite-dimensional degeneracy of the quasinormal modes in the SAdS$_4$ case has become infinite-dimensional. This is reminiscent of Landau modes (see a related discussion in \cite{Drukker:2003mg}).

In our discussion for the solutions of angular equation we were agnostic regarding the possible values of the parameters $q,m$ and $\Omega$. For example, they can take complex values which we denote as
\be
\label{qmOmega}
q=q_1+iq_2\,,\,\,\,m=m_1+im_1\,,\,\,\,\Omega=\Omega_1+i\Omega_2\,.
\ee
Consider then a highest weight state with $q=m$ and use the condition that $m-\Omega={\cal N}\in{\mathbb R}$. These lead to 
\be
\label{qOmega}
q=q_1+iq_2=m_1+im_2=\Omega_1+{\cal N}+i\Omega_2=\Omega+{\cal N}\,.
\ee
This determines the separation constant $C$ of the angular equation \ref{TN2}) as\footnote{If we further require this to be real we find \cite{Kalamakis:2020aaj}
$
C=-\frac{1}{4}-\Omega_2^2\,,\,\,\,\,{\cal N}=-\frac{1}{2}-\Omega_1
$.
}
\be
\label{C}
C=(\Omega+{\cal N}+1)(\Omega+{\cal N})\,.
\ee

\section{Numerical evaluation of the quasinormal modes}

Our deliberations above did not fully address the issue with the angular equation, however they provide a working path towards the numerical calculation of the quasinormal modes. Our strategy is to substitute (\ref{C}) for $C$ in the radial equation (\ref{TN1}). This way ${\cal N}$ remains as an arbitrary real parameter. In the following we will consider the case ${\cal N}=0$ and we will see that the dependence of our results on ${\cal N}\in [-2,2]$ is rather weak. 
\subsection{Qualitative features}
The treatment of the radial equation proceeds in a standard manner. We start by setting  
\be\label{RtoZ}
R(r)=\frac{1}{\sqrt{G(r,n)}}Z(r)\,,
\ee
and we obtain from (\ref{TN1})
\be\label{radialEqZ1}
V(r,n)Z''(r)+V'(r,n) Z'(r)+\left[\omega^2\frac{h^2(r,n)}{V(r,n)}-U(r,n)\right]Z(r)=0\,,
\ee
with
\be 
\label{UTNh}
 U(r,n) =
 \frac{rV'(r,n)+C}{G(r,n)}+\frac{n^2V(r,n)}{G^2(r,n)}\,,\qquad
  h^2(r,n)=1+\frac{4n^2V(r,n)}{G(r,n)}\,,
\ee
 where the prime denotes differentiation with respect to $r$. We can also write $V(r,n)$ in terms of the horizon distance $r_+$ by eliminating M in (\ref{defr+}) as
 \be
 \label{Vr+}
 V(r,n)=\frac{1}{r^2+n^2}\left(r^2-n^2+ \frac{1}{L^2}\left(r^4+6n^2r^2-3n^4\right)-\frac{r}{L^2r_+}\left(r_+^4+L^2r_+^2+6n^2r_+^2-L^2n^2-3n^4\right)\right)\,.
 \ee
$V(r,n)$ is positive for all $r\in[r_+,\infty)$ and it develops a local maximum and a local minimum when $n$ is large enough compared to $r_+$ \cite{Kalamakis:2020aaj}.  The temperature $T$ is defined in the standard way as (we set $L=1$ onwards)
 \be
\label{TNtemp}
T=\frac{V'(r_+,n)}{4\pi}= \frac{1+3r_+^2+3n^2}{4\pi r_+}\,,
\ee
and for fixed $r_+$  it grows as $T\sim n^2$. 

Next we impose the infalling boundary conditions at the horizon writing
\be\label{ZPsi}
Z(r)=e^{-i\omega r_*}\Psi(r)\,.
\ee
The tortoise coordinate $r_*$ is defined as\footnote{In the tortoise coordinate the horizon is at $r_*\rightarrow-\infty$ and the boundary at $r_*\rightarrow\infty$. }
 \be
 \label{r*}
 \frac{dr_*}{dr}=\frac{h(r,n)}{V(r,n)}\,\Rightarrow\,r_*\sim \frac{1}{4\pi Tr_+}\ln(r-r_+)+\cdots\,,
 \ee
with the dots denoting higher powers in $(r-r_+)$. 
This brings (\ref{radialEqZ1}) to the form
\begin{equation}
\label{radialEqPsi}
V(r,n)\Psi''(r)+[V'(r,n)-2i\omega h(r,n)]\Psi'(r)-[i\omega h'(r,n)+U(r,n)]\Psi(r)=0\,.
\end{equation}
This form is useful in order to acquire a qualitative picture for the quasinormal modes. Multiplying (\ref{radialEqPsi}) with $\Psi^*(r)$ and integrating from $r_+$ to infinity we obtain
\begin{equation}
\label{radialEqPsiQNM1}
\int_{r_+}^\infty\left(-V(r,n)|\Psi'(r)|^2-2i\omega h(r,n)\Psi^*(r)\Psi'(r)-[i\omega h'(r,n)+U(r,n)]|\Psi(r)|^2\right)dr=0\,.
\eeq
Taking the complex conjugate of the above and setting  $C=(\Omega+1)\Omega$ we get
\beq
\label{radialEqPsiQNM2}
\int_{r_{+}}^\infty\left[V(r,n)|\Psi'(r)|^2+{\cal A}(r,n;\omega)|\Psi(r)|^2\right]dr=-\frac{|\omega|^2}{{Im(\omega)}}|\Psi(r_+)|^2\,.
\eeq
with
\beq
\label{A}
{\cal A}(r,n;\omega)=\frac{rV'(r,n)}{G(r,n)}+\frac{n^2V(r,n)}{G^2(r,n)}-\frac{4n^2|\omega|^2}{G(r,n)}\,.
\eeq
When $n=0$ the integrand on the  l.h.s. of (\ref{radialEqPsiQNM2}) is positive, which shows that the quasinormal modes of SAdS$_4$ can only have negative imaginary part and hence they are stable. For $n\neq 0$ we note\footnote{${\cal A}(r,n;\omega)$ depends on $n^2$.} that
\beq
\label{Alimits}
\lim_{r\rightarrow r_+}{\cal A}(r,n;\omega)=\frac{1}{r_+^2+n^2}\left(1+3r_+^2+3n^2-4n^2|\omega|^2\right)\,,\,\,\,\lim_{r\rightarrow\infty}{\cal A}(r,n;\omega)=2\,.
\eeq
Therefore, if 
\beq
\label{omegabound}
|\omega|^2\leq |\omega_*(r_+,n)|^2=\frac{1+3r_+^2+3n^2}{4n^2}\,,
\eeq
then ${\cal A}(r,n;\omega)$ is positive for $r\in[r_+,\infty)$ and hence the corresponding quasinormal modes are stable. 
The typical behaviour of ${\cal A}(r,n,;\omega)$ is plotted for $r_+=n=1$  in Fig.1. 
\begin{figure}[h]
\centering
\includegraphics{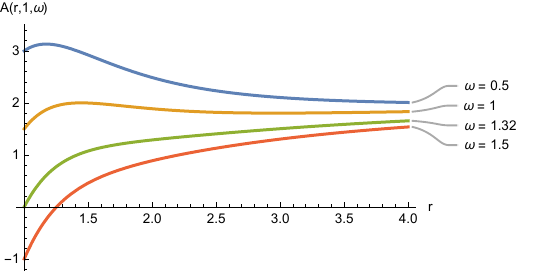}
\caption{Typical plots of $A(r,1;\omega)$ for $|\omega_*|=\frac{\sqrt{7}}{2}\approx 1.32.$}
\end{figure}
This analysis shows that for a given $n\neq 0$ there always exist a region in the complex plane of the quasinormal modes $\omega$, which can be thought of as a circle centered at the origin, that contains only stable modes. The value $|\omega_*(r_+,n)|$ is clearly a lower bound for the radius of this  {\it stable QNM region} and it grows linearly with $r_+$ for large $r_+$. Hence large TNAdS$_4$ black holes have a larger stable QNM region. On the other hand  $|\omega_*(r_+,n)|$ for fixed $r_+$ is proportional to $1/n$  therefore the stable QNM region shrinks as $n$ increases. Interestingly though, this region never shrinks to a point as the limit $n\rightarrow\infty$ of (\ref{omegabound}) is finite and gives
\beq
\label{omegabound34}
\lim_{n\rightarrow\infty}|\omega_*(r_+,n)|=\frac{\sqrt{3}}{2}\,.
\eeq
It would be nice to understand this limiting case further. 

\subsection{Setting up the numerical algorithm}
We set up the numerical algorithm for the calculation of the QNMs using following the seminal work \cite{Horowitz:1999jd}. If we  define
\be\label{Psipsi}
h(r,n)\psi(r)=\Psi(r)\,,
\ee
the radial equation becomes
 \begin{align}
 \label{radialEqPsi1}
 & V(r,n)\psi''(r)+V_1(r,n)\psi'(r) +V_2(r,n)\psi(r)=0\,,
 \end{align}
 where
 \begin{align}
 \label{V1}
 &V_1(r,n)=V'(r,n)-V(r,n)\frac{h'(r,n)}{h(r,n)}-2i\omega h(r,n)\,,\\
 \label{V2}
 &V_2(r,n)=\frac{3V(r,n)(h'(r,n))^2}{4h(r,n)^2}-\frac{V(r,n)h''(r,n)}{2h(r,n)}-\frac{V'(r,n)h'(r,n)}{2h(r,n)}-U(r,n)\,.
 \end{align}
We now ask that $\psi(r_+)$ is finite and as a final step we make the change of variables $z=r_+/r$ such that the horizon is  at $z=1$ and the boundary at $z=0$. In the $z$ variable the potential $V(r,n)\mapsto V(z,n)$ with
 \be
 \label{Vz}
V(z,n)=\frac{\frac{r_+^4}{z^2}+(1+6n^2)r_+^2-z\left[r_+^2(1+r_+^2)-n^2(1-6r_+^2+3n^2)\right]-n^2(1+3n^2)z^2}{r_+^2+n^2z^2}\,,
 \ee
and the final form of the radial equation becomes
\be\label{radialEqPsi2}
s(z)\psi''(z)+\frac{t(z)}{z-1}\psi'(z)-\frac{u(z)}{(z-1)^2}\psi(z)=0\,,
\ee
% \beq
% \psi'(r)=\frac{dz}{dr}\psi'(z)=-\frac{r_+}{r^2}\psi'(z)=-\frac{z^2}{r_+}\psi'(z)
% \eeq
%\beq
%\psi''(r)=-\frac{dz}{dr}\Big(\frac{2z}{r_+}\psi'(z)+\frac{z^2}{r_+}\psi''(z)\Big)
%=\frac{z^3}{r_+^2}\Big(2\psi'(z)+z\psi''(z)\Big)
%\eeq
%\begin{align}
%& V(z)\Psi''(z)
%+\frac{r_+}{z^2}\Big(\frac{2zV(z)}{r_+}-V_1(z)\Big)\Psi'(z) 
%+\frac{r_+^2}{z^4}V_2(z)\Psi(z)=0\,,
%\end{align}
%\begin{align}
%&\frac{r_+}{z^2}\Big(\frac{2zV(z)}{r_+}-V_1(z)\Big)
%=\frac{1}{z}\Big(
%2V(z)
%+zV'(z)
%-zV(z)h_1(z)
%+2i \frac{\omega r_+}{z}h(z)
%\Big)
%\,,\\
% &-\frac{r_+^2}{z^4}V_2(z)=U_{TN}(z)+\frac14\Big(
%-3V(z)h_1(z)^2
% +\frac{4V(z)}{z}h_1(z)
% +2h(z)V(z)h_2(z)
% +2V'(z)h_1(z)
% \Big)
% \end{align}
where
\beqn
\label{st}
s(z)&=&V(z,n),\qquad
t(z)=\frac{z-1}{z}\left[2V(z,n)+zV'(z,n)-zV(z,n)h_1(z,n)+\frac{2i\omega r_+}{z}h(z,n)\right]\,,\nonumber \\
\label{uz}
u(z)&=&(z-1)^2\left[U(z,n)+\frac{z^4}{4}V(z,n)\Big(-3h_1^2(z,n)+2h(z,n)h_2(z,n)-\frac{4}{z}h_1(z,n)\Big)\right. \nonumber \\ &&\left.\hspace{1.5cm}+\frac{z^4}{2} h_1(z,n)V'(z,n)\right], \nonumber \\
\label{UTNz}
U(z,n)&=& \frac{C z^2}{r_+^2+n^2z^2}-\frac{z^3}{r_+^2+n^2z^2}V'(z,n)+\frac{z^4 n^2V(z,n)}{(r_+^2+n^2z^2)^2}\,,\nonumber 
\eeqn
%
% where
%\begin{align}
%\label{st}
%&\hspace{-.1cm}s(z)=\frac{1}{r_+^2}V(z),\qquad
%t(z)=\frac{z-1}{z\,r_+^2}\left[2V(z)+zV'(z)-zV(z)h_1(z)+\frac{2i\omega r_+}{z}h(z)\right]\,,\\
%\label{uz}
%&\hspace{-.6cm}u(z)=\frac{(z-1)^2}{z^4}\left[U_{TN}(z)+\frac{V(z)\left[-3z^4h_1(z)^2+2z^4h(z)h_2(z)+4z^3h_1(z)\right]+ 2z^4 h_1(z)V'(z)}{4r_+^2}\right], \\
% \label{UTNz}
%& \hspace{-.1cm}U_{TN}(z)= m_\phi^2+\frac{\lambda z^2}{r_+^2+n^2z^2}-\frac{z^3}{r_+^2+n^2z^2}V'(z)+\frac{z^4 n^2V(z)}{(r_+^2+n^2z^2)^2}\,
% \end{align}
and we have defined
 \be
 \label{h1h2}
 h_1(z,n)=\frac{h'(z,n)}{h(z,n)},\qquad h_2(z,n)=\frac{h''(z,n)}{h^2(z,n)}.
 \ee
One can check that for $n=0$ the functions above coincide\footnote{When $n=0$ we have to set $\Omega=0$ and ${\cal N}=m\in{\mathbb Z}$.} with the ones in \cite{Horowitz:1999jd}. Now the horizon $z=1$ is a regular singular point of the differential equation (\ref{radialEqPsi2}) since the functions $s(z)$, $t(z)$ and $u(z)$ have regular expansions around it as
 \be
 \label{stu}
 s(z)=\sum_{k=0}^{\infty}s_k(z-1)^k\,,\,\,\,t(z)=\sum_{k=0}^{\infty}t_k(z-1)^k\,,\,\,\,u(z)=\sum_{k=0}^{\infty}u_k(z-1)^k\,,
 \ee
Therefore, near $z=1$ we  look for a series solution of the form
\be\label{Psiseries}
\psi(z)=(z-1)^a\sum_{k=0}^\infty \psi_k(\omega)(z-1)^k\,,
\ee
%\be
%\Psi'(z)=\sum_{k=1}^\infty k\Psi_k(\omega)(z-1)^{k-1}%=\sum_{j=0}^\infty (j+1)\Psi_{j+1}(\omega)(z-1)^{j}
%\ee
%\be
%\Psi''(z)=\sum_{k=2}^\infty k(k-1)\Psi_k(\omega)(z-1)^{k-2}%=\sum_{j=0}^\infty (j+1)(j+2)\Psi_{j+2}(\omega)(z-1)^{j}
%\ee
%
%\beq
%\sum_{i=0,j=2}^{\infty}s_ij(j-1)\Psi_{j}(\omega)(z-1)^{i+j-2}
%+\sum_{i=0,j=1}^{\infty}t_ij\Psi_{j}(\omega)(z-1)^{i+j-2}
%-\sum_{i=1,j=0}^{\infty}u_i\Psi_{j}(\omega)(z-1)^{i+j-2}
%=0
%\eeq
%\beq
%t_0\Psi_{1}(\omega)
%=u_1\Psi_{0}(\omega)
%\eeq
%\beq
%(2s_0+2t_0)\Psi_{2}(\omega)
%=u_2\Psi_{0}(\omega)
%+(u_1-t_1)\Psi_{1}(\omega)
%\eeq
%\beq
%(6s_0+3t_0)\Psi_{3}(\omega)
%=(u_1-2s_1-2t_1)\Psi_{2}(\omega)
%+(u_2-t_2)\Psi_{1}(\omega)
%+u_3\Psi_{0}(\omega)
%\eeq
where the coefficients $\psi_k$ are functions of $\omega$ and ${\cal N}$ through the relation $C=C(\Omega,{\cal N})$ in (\ref{C}). 

From (\ref{radialEqPsi2}) using (\ref{Psiseries}) we find the indicial equation at the horizon as
 \be
 \label{indicial}
 a(a-1)s_0+at_0-u_0=0\,,
 \ee
where
 \be
 \label{s0t0u0}
 s_0=-\frac{4\pi T}{r_+}\,,\,\,\,t_0=-\frac{2}{r_+}(2\pi T-i\omega)\,,\,\,\,u_0=0\,,
 \ee
and the temperature is defined in (\ref{TNtemp}). The indicial equation is solved giving $a=0$ or $a=i\omega/2\pi T$. The second solution would give the outgoing modes near the horizon, so we set $a=0$ from now on. 
 
Substituting (\ref{stu}) and (\ref{Psiseries}) into (\ref{radialEqPsi2}) we get the following recursion relation for the coefficients $\psi_m(\omega)$
 \be
 \label{recursion}
 \psi_m(\omega)=\frac{1}{m(m-1)s_0+mt_0}\sum_{k=0}^{m-1}\Big[u_{m-k}-kt_{m-k}-k(k-1)s_{m-k}\Big]\psi_k(\omega)\,,
 \ee
 Finally, the quasinormal modes are the complex roots $\omega$ of the polynomial equation that enforces the Dirichlet boundary conditions at the boundary, namely
 \be
 \label{eqhorizon}
 \psi(0)=\sum_{k=0}^\infty\psi_k(\omega)(-1)^k=0\,.
 \ee

\subsection{Numerical results and critical NUT charge}
Our numerical routines follow closely the standard ones\footnote{See for example https://centra.tecnico.ulisboa.pt/network/grit/files/ringdown/.}  and can be made available upon request. From the outset it is important to emphasise that below we present results for "half" of the quasinormal modes, namely those whose lowest mode has a positive real part. The other half can be obtained by the transformation $\omega \rightarrow -\omega^*$ which is a symmetry of the equations as explained above. For clarity and numerical stability we focus on two regimes:  a) "large" BHs with $r_+\gg 1$ and b) "intermediate" BHs with $r_+\sim 1$. A salient feature of the SAdS$_4$ QNMs is that the ratio $\omega/r_+$ for the lowest QNM is constant for both large and intermediate BHs, and it is only slightly different for the two regimes. The proportionality constant depends on the angular momentum of the modes.

Firstly we focus on the lowest QNMs. We present in Tables 1 and 2 below the numerical results for the real part $\omega_R$ and for minus the imaginary part $-\omega_I$ of the lowest QNMs, as well and for its norm $|\omega|$, for typical values of $r_+$ in the large and intermediate regimes. We start our calculations for $n=0$ to reproduce the results of \cite{Horowitz:1999jd} and then we increase $n$. 

\begin{table}[!h]
\centering
\caption{The lowest QNMs in intermediate TNAdS$_4$ black holes}
\begin{tabular}{|ccccc|}
\hline
 $r_+$ & $n$ & Re[$\omega$] & -Im[$\omega$] & $|\omega|$ \\
\hline
0.4 & 0.000 & 2.37 & 1.16 & 2.64 \\
0.4 & 0.001 & 2.36 & 1.16 & 2.63 \\
0.4 & 0.002 & 2.36 & 1.16 & 2.63 \\
0.4 & 0.01 & 2.34 & 1.15 & 2.61 \\
0.4 & 0.1 & 2.13 & 1.13 & 2.41 \\
0.6 & 0.000 & 2.45 & 1.58 & 2.91 \\
0.6 & 0.001 & 2.44 & 1.58 & 2.91 \\
0.6 & 0.002 & 2.44 & 1.58 & 2.91 \\
0.6 & 0.01 & 2.42 & 1.57 & 2.89 \\
0.6 & 0.1 & 2.23 & 1.55 & 2.72 \\
\hline
\end{tabular}
\qquad
\begin{tabular}{|ccccc|}
\hline
 $r_+$ & $n$ & Re[$\omega$] & -Im[$\omega$] & $|\omega|$ \\
\hline
0.8 & 0.000 & 2.59 & 2.13 & 3.35 \\
0.8 & 0.001 & 2.59 & 2.13 & 3.35 \\
0.8 & 0.002 & 2.58 & 2.13 & 3.35 \\
0.8 & 0.01 & 2.57 & 2.12 & 3.33\\
0.8 & 0.1 & 2.39 & 2.06 & 3.15 \\
1.0 & 0.000 & 2.8 & 2.67 & 3.87 \\
1.0 & 0.001 & 2.8 & 2.67 & 3.87 \\
1.0 & 0.002 & 2.79 & 2.67 & 3.86 \\
1.0 & 0.01 & 2.78 & 2.66 & 3.84 \\
1.0 & 0.1 & 2.6 & 2.58 & 3.66 \\
\hline
\end{tabular}
\label{Table2}
\end{table}

\begin{table}[!h]
\centering
\caption{The lowest QNMs in large TNAdS$_4$ black holes}
\begin{tabular}{|ccccc|}
\hline
 $r_+$ & $n$ & Re[$\omega$] & -Im[$\omega$] & $|\omega|$ \\
\hline
 5 & 0.000 & 9.47 & 13.33 & 16.35 \\
 5 & 0.001 & 9.47 & 13.32 & 16.35 \\
 5 & 0.002 & 9.47 & 13.32 & 16.34 \\
 5 & 0.010 & 9.45 & 13.31 & 16.32 \\
 5 & 0.100 & 9.27 & 13.16 & 16.09 \\
10 & 0.000 & 18.61 & 26.64 & 32.5 \\
10 & 0.001 & 18.6 & 26.64 & 32.49 \\
10 & 0.002 & 18.6 & 26.64 & 32.49 \\
10 & 0.01 & 18.59 & 26.62 & 32.47 \\
10 & 0.1 & 18.41 & 26.47 & 32.24 \\
 \hline
\end{tabular}
\qquad
\begin{tabular}{|ccccc|}
\hline
 $r_+$ & $n$ & Re[$\omega$] & -Im[$\omega$] & $|\omega|$ \\
\hline
 50 & 0.000 & 92.49 & 133.19 & 162.16 \\
 50 & 0.001 & 92.49 & 133.19 & 162.16 \\
 50 & 0.002 & 92.49 & 133.19 & 162.15 \\
 50 & 0.01 & 92.47 & 133.18 & 162.13 \\
 50 & 0.1 & 92.33 & 133.03 & 161.93 \\
100 & 0.000 & 184.95 & 266.38 & 324.3 \\
100 & 0.001 & 184.95 & 266.38 & 324.29 \\
100 & 0.002 & 184.95 & 266.38 & 324.29 \\
100 & 0.01 & 184.93 & 266.37 & 324.27 \\
100 & 0.1 & 184.84 & 266.24 & 324.11 \\
\hline
\end{tabular}
\label{Table1}
\end{table}

For small values of $n\in[0,0.1]$, we notice that $\omega_R,\omega_I$ and $|\omega|$ they all decrease but only  slightly w.r.t the SAdS$_4$ values. 
%For these values the ratio $\omega/r_+$ stays constant in both regimes and decreases slightly. 
In the large black hole regime the ratio $|\omega|/r_+$ stays approximately constant, while in the intermediate regime the relation exhibits a mild non-linearity. This can be seen in figures \ref{fig:lowest_qnm_regression}. The approximate formulae for the large and intermediate regime are
\beq
\label{omegar+}
%|\omega|_{large}=C_1r_+\quad|\omega|_{small}=C_2r_+\,.
%|\omega|_{intermediate}=1.51r^2_+ + 2.41\quad|\omega|_{large}=3.24r_+\,.
|\omega|_{intermediate}=2.07r_+ + 1.74\quad|\omega|_{large}=3.24r_+\,.
\eeq

\begin{figure}[!h]%
    \centering
    \subfigure[Intermediate horizon regime.]{\includegraphics[width=0.49\textwidth]{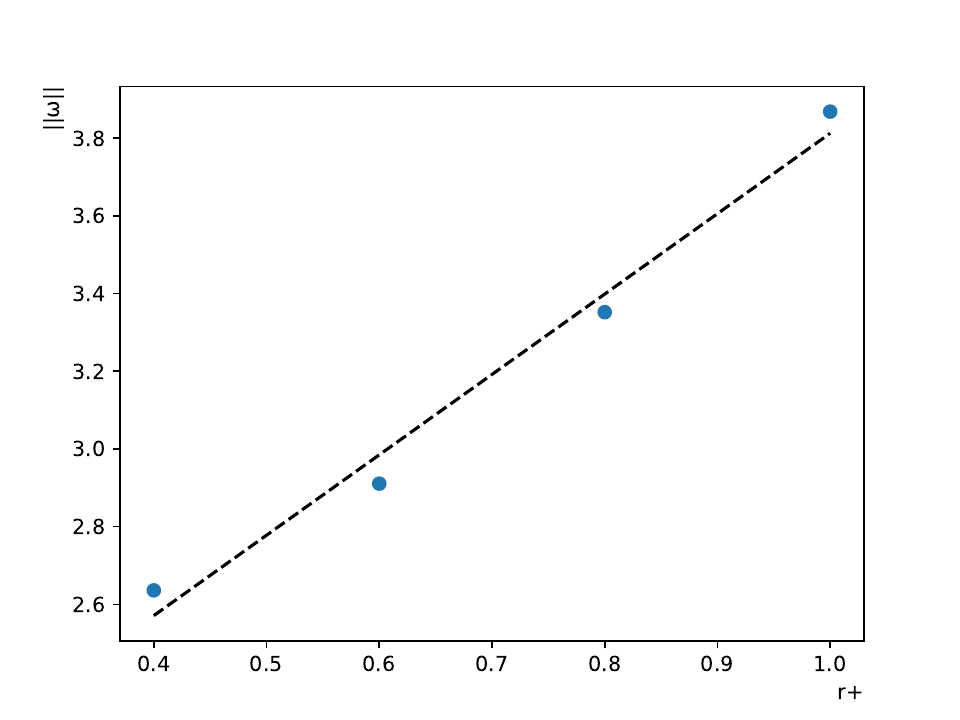}}
    \subfigure[Large horizon regime]{\includegraphics[width=0.49\textwidth]{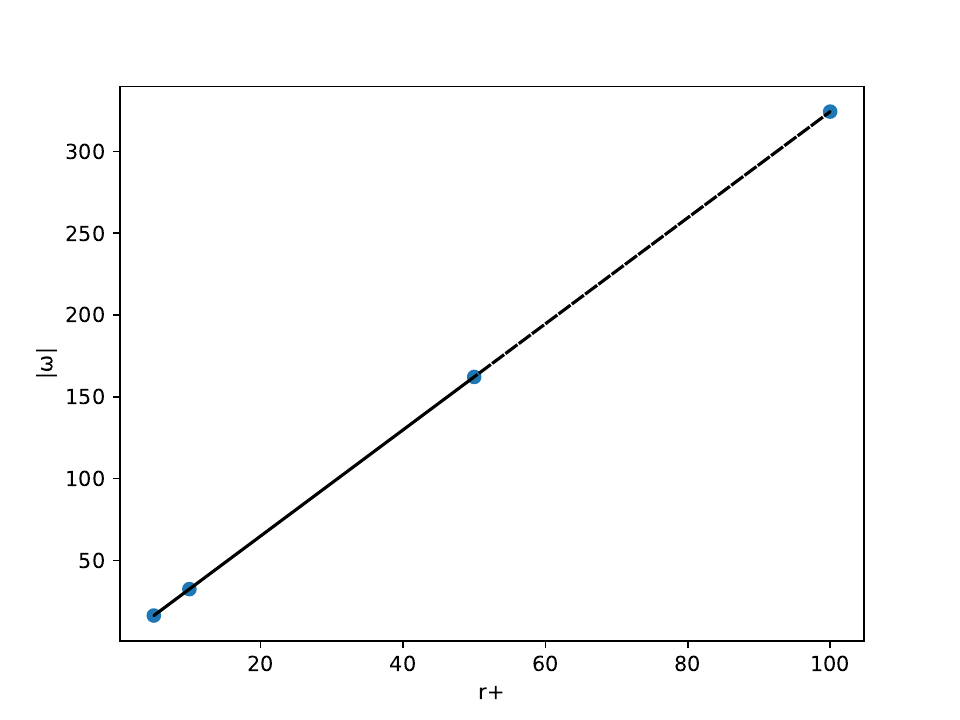}}
    \caption{Lowest QNM regression wrt the black hole horizon.}
    \label{fig:lowest_qnm_regression}
\end{figure}
   
Remarkably, when $n$ approaches a critical value $n_{cr}$ the lowest QNMs become overdamped as their real part drops abruptly towards zero.  On the other hand, their imaginary part drops only slightly. This behaviour of $\omega_R,\omega_I$ and $|\omega|$ is depicted in figures \ref{fig:qnm_crit_intermediate}, \ref{fig:qnm_crit_large}.

\begin{table}[!h]
\centering
\caption{The lowest QNMs of intermediate and large TNAdS$_4$ black holes for values of the NUT charge close to and beyond the critical value.}
\begin{tabular}{|ccccc|}
\hline
 $r_+$ & $n$ & Re[$\omega$] & -Im[$\omega$] & $|\omega|$ \\
\hline
0.4 & 0.2 & 0.06 & 2.08 & 2.08 \\
0.4 & 0.25 & 0.06 & 2 & 2 \\
0.4 & 0.3 & 0.05 & 2.14 & 2.14 \\
0.4 & 0.5 & 0.08 & 2.72 & 2.73 \\
0.4 & 1 & 0.04 & 5.58 & 5.58 \\
0.4 & 5 & 0.01 & 95.6 & 95.6 \\
0.6 & 0.2 & 0.08 & 0.45 & 0.46 \\
0.6 & 0.25 & 0.15 & 1.32 & 1.33 \\
0.6 & 0.3 & 0.17 & 1.44 & 1.45 \\
0.6 & 0.5 & 0.22 & 1.99 & 2 \\
0.6 & 1 & 0.2 & 4.08 & 4.09 \\
0.6 & 5 & 0.05 & 64.23 & 64.23 \\
0.8 & 0.2 & 2.17 & 2.22 & 3.1 \\
0.8 & 0.25 & 0.12 & 1.35 & 1.36 \\
0.8 & 0.3 & 0.13 & 1.49 & 1.49 \\
0.8 & 0.5 & 0.15 & 1.98 & 1.98 \\
0.8 & 1 & 0.13 & 3.59 & 3.59 \\
0.8 & 5 & 0.03 & 48.7 & 48.7 \\
1 & 0.2 & 2.45 & 2.55 & 3.53 \\
1 & 0.25 & 0.11 & 1.35 & 1.36 \\
1 & 0.3 & 0.12 & 1.57 & 1.57 \\
1 & 0.5 & 0.14 & 2.03 & 2.04 \\
1 & 1 & 0.11 & 3.38 & 3.38 \\
1 & 5 & 0.03 & 39.49 & 39.49 \\
\hline
\end{tabular}
\qquad
\begin{tabular}{|ccccc|}
\hline
 $r_+$ & $n$ & Re[$\omega$] & -Im[$\omega$] & $|\omega|$ \\
\hline
5 & 0.2 & 9.15 & 13.01 & 15.91 \\
5 & 0.25 & 8.26 & 13.7 & 16 \\
5 & 0.3 & 0.1 & 5.07 & 5.07 \\
5 & 0.5 & 0.11 & 6.29 & 6.29 \\
5 & 1 & 0.090 & 7.41 & 7.41 \\
5 & 5 & 0.02 & 15.08 & 15.08 \\
10 & 0.2 & 18.35 & 26.33 & 32.1 \\
10 & 0.25 & 16.94 & 27.49 & 32.29 \\
10 & 0.3 & 0.1 & 9.91 & 9.91 \\
10 & 0.5 & 0.11 & 12.34 & 12.34 \\
10 & 1 & 0.09 & 14.22 & 14.22 \\
10 & 5 & 0.02 & 18.75 & 18.75 \\
50 & 0.2 & 92.88 & 133 & 162.22 \\
50 & 0.25 & 86.59 & 138.53 & 163.36 \\
50 & 0.3 & 0.1 & 49.16 & 49.16 \\
50 & 0.5 & 0.11 & 61.29 & 61.29 \\
50 & 1 & 0.09 & 70.15 & 70.15 \\
50 & 5 & 0.02 & 75.53 & 75.53 \\
100 & 0.2 & 186.16 & 266.35 & 324.96 \\
%100 & 0.25 & 173.69 & 277.39 & 327.28 \\
100 & 0.25 & 155.05 & 288.64 & 327.65 \\
100 & 0.3 & 0.10 & 98.3 & 98.3 \\
100 & 0.5 & 0.11 & 122.55 & 122.55 \\
100 & 1 & 0.09 & 140.25 & 140.25 \\       % Is unstable at Nit = 30
100 & 5 & 0.02 & 149.92 & 149.92 \\
\hline
\end{tabular}
\label{Table3}
\end{table}

The behaviour of the lowest QNM for values of the NUT charge close to and beyond the critical limit is shown in tables \ref{Table3}.

%\begin{figure}[!h]
%\centering
%\includegraphics[width=0.85\textwidth]{qnm_crit_large.pdf}
%\caption{Overdamping of the lowest QNM at the critical Nut charge limit for large black holes.}
%\end{figure}

\begin{figure}[h]%
    \centering
    \subfigure[Real part of the lowest QNM.]{\includegraphics[width=0.49\textwidth]{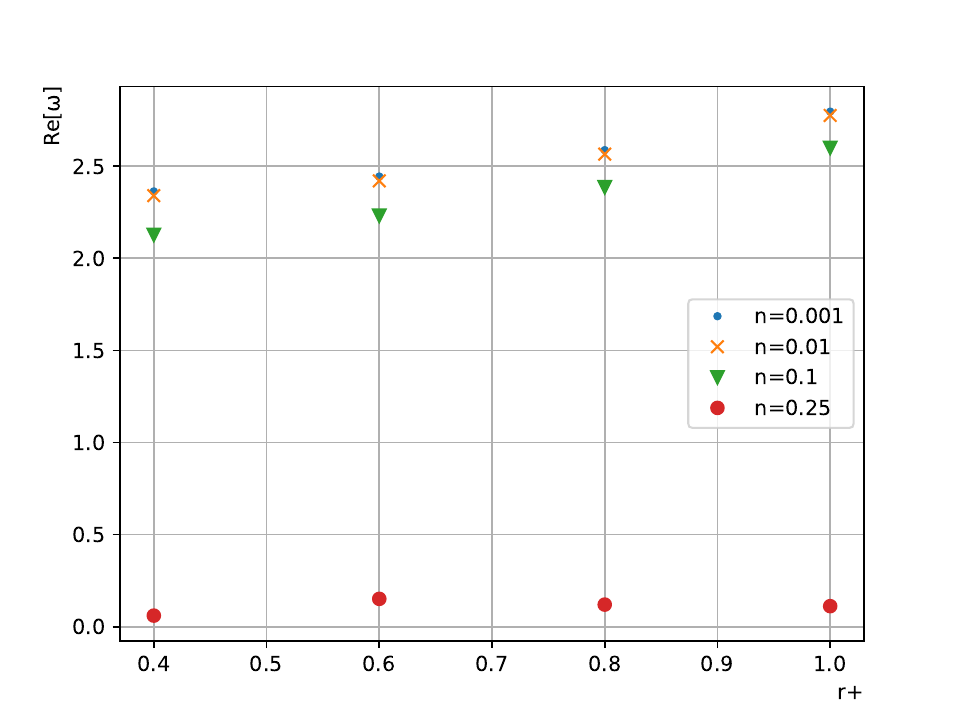}}
    \subfigure[Imaginary part of the lowest QNM.]{\includegraphics[width=0.49\textwidth]{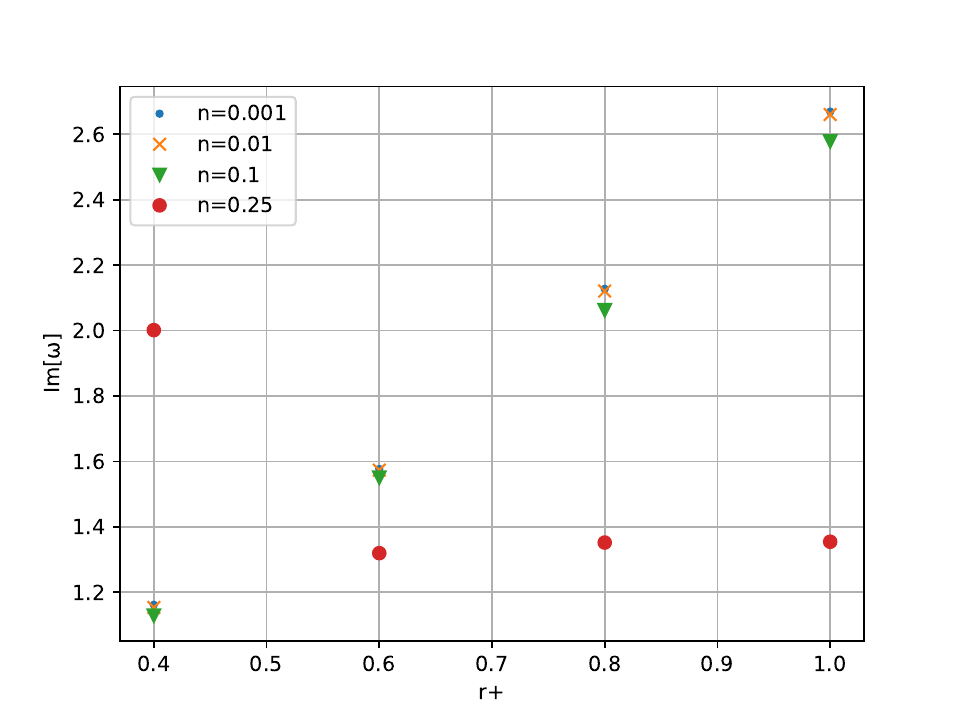}}
    \caption{Overdamping of the lowest QNM at the critical Nut charge limit for intermediate black holes.}
    \label{fig:qnm_crit_intermediate}
\end{figure}

\begin{figure}[h]%
    \centering
    \subfigure[Real part of the lowest QNM.]{\includegraphics[width=0.49\textwidth]{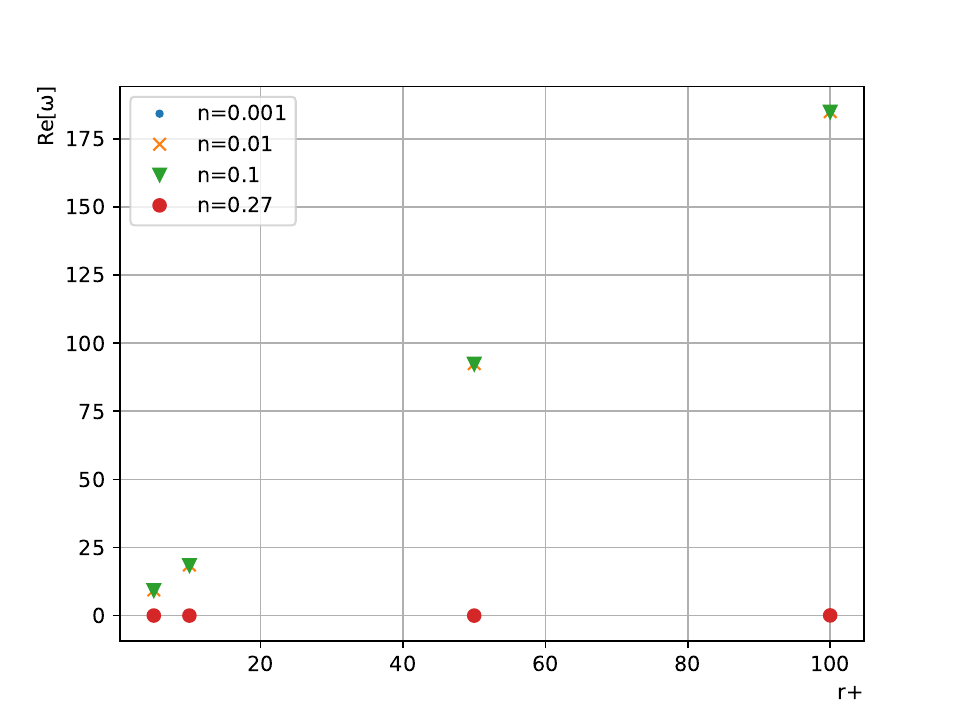}}
    \subfigure[Imaginary part of the lowest QNM.]{\includegraphics[width=0.49\textwidth]{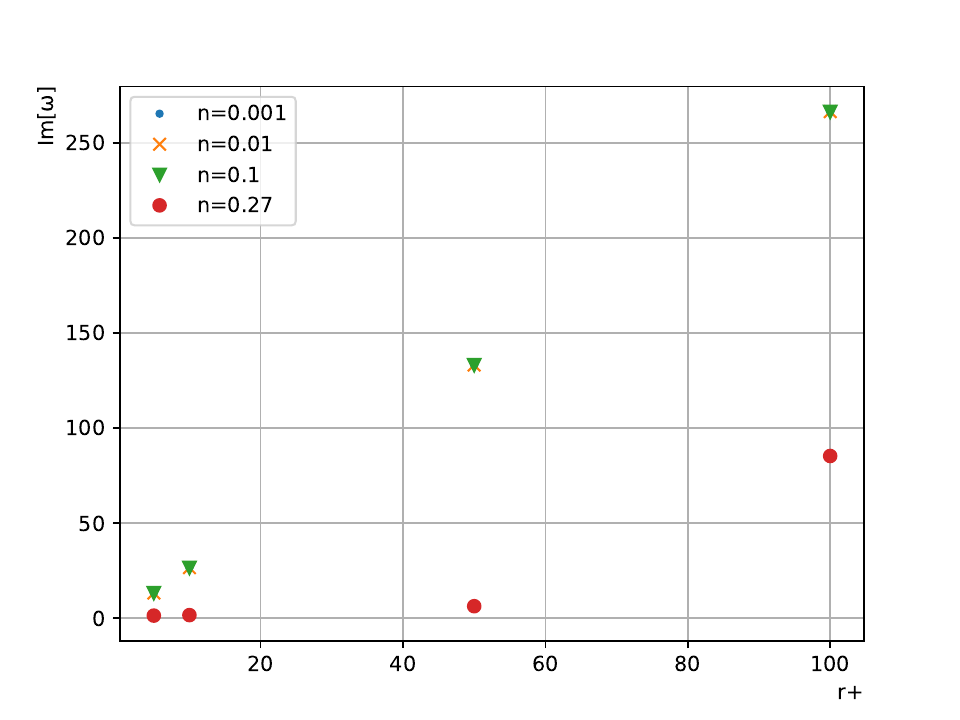}}
    \caption{Overdamping of the lowest QNM at the critical Nut charge limit for large black holes.}
    \label{fig:qnm_crit_large}
\end{figure}

Our numerical results show that the critical value $n_{cr}$ where this overdamping takes place is almost the same for the whole range of $r_+$ that we considered. This begs an analytic explanation that we will now attempt. For a given $r_+$ and $n$ we have introduced the bound $|\omega_*(r_+,n)|$ in (\ref{omegabound}) for the stable QNM region. But according to our results the norm of the lowest stable QNM is related to $r_+$ for both the large and intermediate regimes as per the figures \ref{fig:lowest_qnm_regression}. The approximate relations for these regimes are shown in (\ref{omegar+}).  Now as $n$ increases, for a given $r_+$, the bound $|\omega_*(r_+,n)|$ decreases and at some point it will become equal and then smaller than the norms (\ref{omegar+}). So equating the $|\omega_*(r_+,n)|$ with the norms (\ref{omegar+}) we will find the value $n_{cr}$ beyond which the radius of the stable QNM region is smaller that the radius of the lowest QNM for a given $r_+$. Assuming a generic behaviour $|\omega|=ar_++b$ we find
\beq
\label{ncrit}
(ar_++b)^2=\frac{1+3r_+^2+3n_{cr}^2}{4n_{cr}^2}\quad\Rightarrow \quad n_{cr}=\sqrt{\frac{1+3r_+^2}{4a^2r^2_++4b^2+8abr_+-3}}\,.
\eeq
\begin{figure}[h]%
    \centering
    \subfigure[Intermediate horizon regime.]%{\includegraphics[width=0.49\textwidth]{n_crit_intermediate.png}}
{\includegraphics[width=0.49\textwidth]{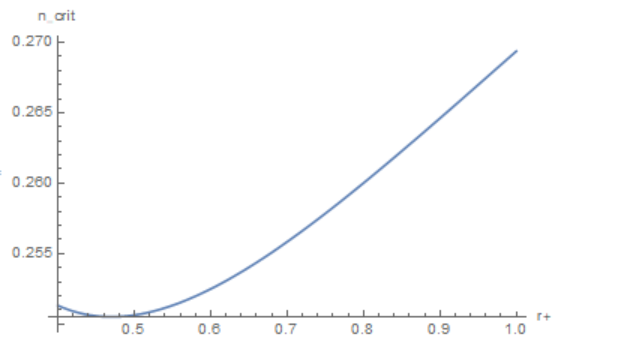}}
    \subfigure[Large horizon regime.]{\includegraphics[width=0.49\textwidth]{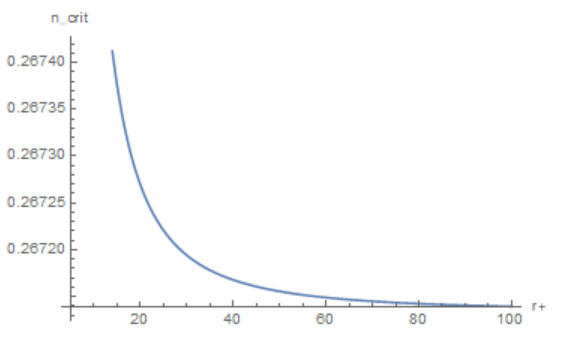}}
    \caption{Behaviour of the critical NUT charge for different horizons.}
    \label{fig:n_critical}
\end{figure}
Substituting then $a=3.24, b=0$ for large BHs, and $a=2.07, b=1,74$ for intermediate BHs we can plot the rhs of (\ref{ncrit})  in figures \ref{fig:n_critical} we see that for  $r_+\in[0.4,100]$ we have $n_{cr}\in[0.25,0.27]$. These are the values of the NUT charge for which we numerically observe the overdamping of the lowest stable QNMs. Clearly the above quantitative argument is not an analytic proof of our numerical results for the existence of $n_{cr}$. Nevertheless we believe that it shows the physical importance of the stable QNM region for the TNAdS$_4$ black holes.

In figures \ref{fig:w_norm_vs_n} we plot the behaviour of the norm of the lowest mode for different values of the NUT charge in the large and intermediate BHs regimes.  We notice that $|\omega|$ drops significantly for $n\approx n_{cr}$ but it starts increasing for larger values of $n$. 
\begin{figure}[h]%
    \centering
    \subfigure[$r_+=1$]%{\includegraphics[width=0.49\textwidth]{n_crit_intermediate.png}}
{\includegraphics[width=0.49\textwidth]{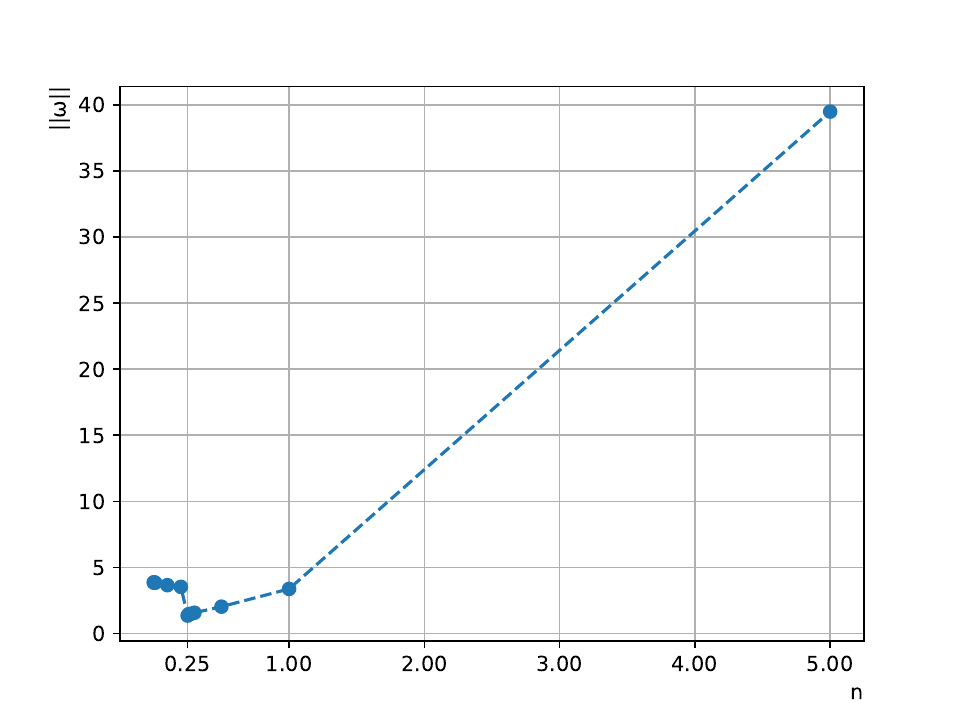}}
    \subfigure[$r_+=100$]{\includegraphics[width=0.49\textwidth]{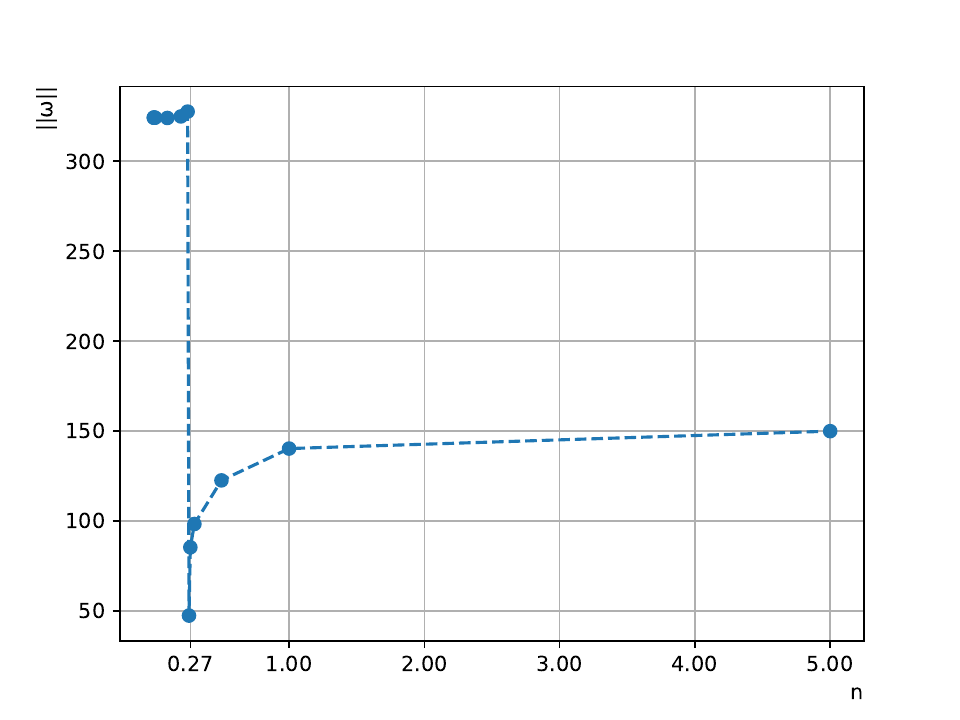}}
    \caption{$|\omega|$ of the lowest QNM for different values of the NUT charge.}
    \label{fig:w_norm_vs_n}
\end{figure}
The imaginary part of the lowest QNM increases sharply towards zero for values of NUT charge that approach the critical n. This is demonstrated in figures \ref{fig:w_im_around_n_crit} for indicative values of the horizon. For larger values of the NUT charge it starts decreasing towards negative values. 

\begin{figure}[h]%
    \centering
    \subfigure[$r_+=1$]%{\includegraphics[width=0.49\textwidth]{n_crit_intermediate.png}}
{\includegraphics[width=0.49\textwidth]{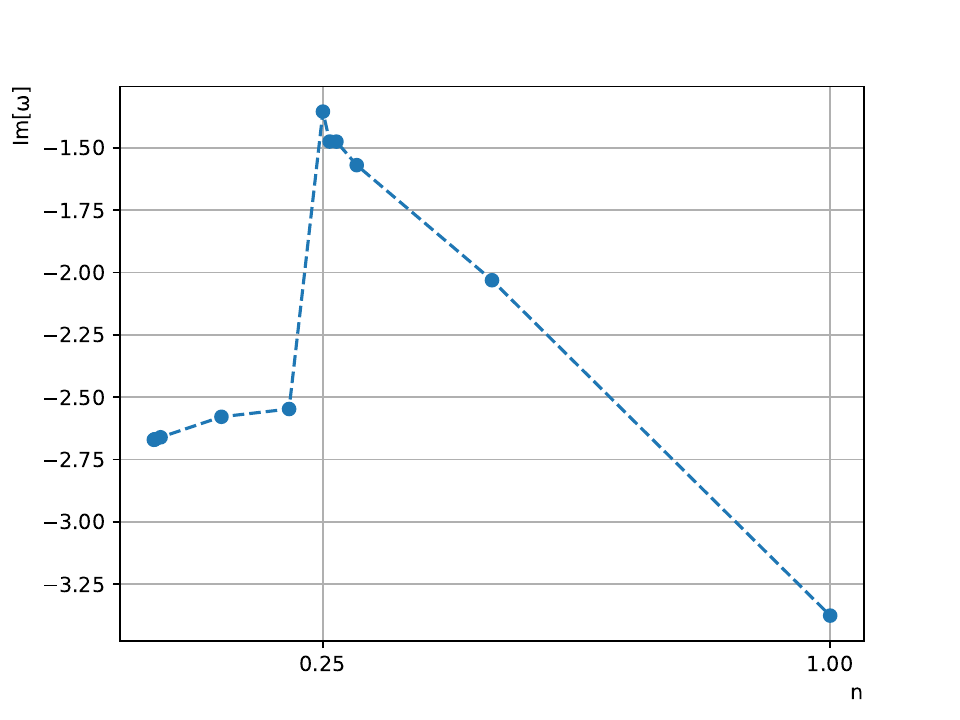}}
    \subfigure[$r_+=100$]{\includegraphics[width=0.49\textwidth]{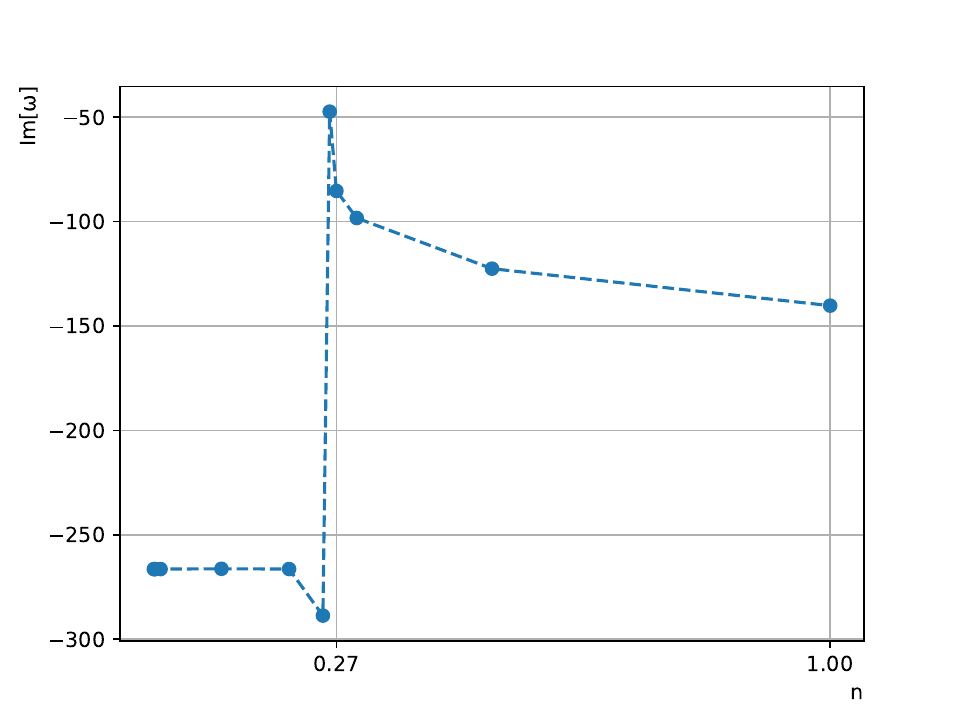}}
    \caption{Imaginary part of the lowest QNM for different values of the NUT charge close to the critical value.}
    \label{fig:w_im_around_n_crit}
\end{figure}

%\clearpage

We may further query how the NUT charge affect the higher overtones. For that we give in the following tables \ref{table:overtones1}-\ref{Table:overtones4} the relevant results for the first ten overtones for the values $r_+=100,1$ and various values of $n$ up to $n_{cr}$. We notice that the presence of a non zero $n$ has introduced a splitting of the two dual QNM modes as $[\omega, -\omega^*]\mapsto [(\omega_+,\omega_-),(-\omega_-^*,-\omega_+^*)]$ where we have denoted with $\omega_+$ the mode with the larger positive real part. The difference $\Delta\omega=\omega_+-\omega_-$ depends on $n$ and appears to be maximum near $n_{cr}$. The study of this interesting property of the QNMs is beyond the scope of the present work, but we hope to return to this issue in the near future. 
\begin{table}[!h]
\centering
\caption{The first 10 quasinormal modes for $r_+=1$, for various values of $n$.}
\begin{tabular}{|cccc|}
\hline
$n$ & $k$ & Re$[\omega]$ & Im$[\omega]$ \\
\hline
0 & 1 & 2.8 & -2.67 \\
0 & 2 & -2.8 & -2.67 \\
0 & 3 & 4.77 & -5.11 \\
0 & 4 & -4.77 & -5.11 \\
0 & 5 & 6.2 & -6.91 \\
0 & 6 & -6.2 & -6.91 \\
0 & 7 & 9.81 & -8.14 \\
0 & 8 & -9.81 & -8.14 \\
0 & 9 & 14.39 & -11.3 \\
0 & 10 & -14.39 & -11.3 \\
\hline
\end{tabular}
\qquad
\begin{tabular}{|cccc|}
\hline
$n$ & $k$ & Re$[\omega]$ & Im$[\omega]$ \\
\hline
0.01 & 1 & 2.78 & -2.66 \\
0.01 & 2 & -2.82 & -2.68 \\
0.01 & 3 & 4.74 & -5.09 \\
0.01 & 4 & -4.8 & -5.12 \\
0.01 & 5 & 6.17 & -6.91 \\
0.01 & 6 & -6.23 & -6.91 \\
0.01 & 7 & 9.74 & -8.13 \\
0.01 & 8 & -9.88 & -8.14 \\
0.01 & 9 & 14.29 & -11.27 \\
0.01 & 10 & -14.5 & -11.34 \\
\hline
\end{tabular}
\label{table:overtones1}
\end{table}

\begin{table}[!h]
\centering
\caption{The first 10 quasinormal modes for $r_+=1$, for various values of $n$.}
\begin{tabular}{|cccc|}
\hline
$n$ & $k$ & Re$[\omega]$ & Im$[\omega]$ \\
\hline
0.1 & 1 & 2.6 & -2.58 \\
0.1 & 2 & -3.07 & -2.78 \\
0.1 & 3 & 4.51 & -4.96 \\
0.1 & 4 & -5.11 & -5.34 \\
0.1 & 5 & 5.96 & -6.85 \\
0.1 & 6 & -6.62 & -6.79 \\
0.1 & 7 & 9.36 & -8.1 \\
0.1 & 8 & -10.89 & -8.21 \\
0.1 & 9 & 13.78 & -11.21 \\
0.1 & 10 & -16.02 & -11.92 \\
\hline
\end{tabular}
\qquad
\begin{tabular}{|cccc|}
\hline
$n$ & $k$ & Re$[\omega]$ & Im$[\omega]$ \\
\hline
0.27 & 1 & -0.11 & -1.47 \\
0.27 & 2 & 0.15 & -4.18 \\
0.27 & 3 & -0.85 & -5.07 \\
0.27 & 4 & 2. & -6.85 \\
0.27 & 5 & -3.28 & -7.53 \\
0.27 & 6 & 4.11 & -9.14 \\
0.27 & 7 & -6.12 & -9.88 \\
0.27 & 8 & 6.78 & -11.73 \\
0.27 & 9 & -9.99 & -12.93 \\
0.27 & 10 & 9.85 & -15.24 \\
\hline
\end{tabular}
\label{table:overtones2}
\end{table}

\begin{table}[!h]
\label{Table:overtones3}
\centering
\caption{The first 10 quasinormal modes for $r_+=100$, for various values of $n$.}
\begin{tabular}{|cccc|}
\hline
$n$ & $k$ & Re$[\omega]$ & Im$[\omega]$ \\
\hline
0 & 1 &   184.95 &  -266.39 \\
0 & 2 &  -184.95 &  -266.39 \\
0 & 3 &   316.15 &  -491.64 \\
0 & 4 &  -316.15 &  -491.64 \\
0 & 5 &   446.73 &  -717.16 \\
0 & 6 & - 446.73 &  -717.16 \\
0 & 7 &   566.19 &  -943.82 \\
0 & 8 &  -566.19 &  -943.82 \\
0 & 9 &   705.97 & -1110.41 \\
0 & 10 & -705.97 & -1110.41 \\
\hline
\end{tabular}
\qquad
\begin{tabular}{|cccc|}
\hline
$n$ & $k$ & Re$[\omega]$ & Im$[\omega]$ \\
\hline
0.01 & 1 &   184.93 &  -266.37 \\
0.01 & 2 &  -184.98 &  -266.40 \\
0.01 & 3 &   316.12 &  -491.61 \\
0.01 & 4 &  -316.18 &  -491.66 \\
0.01 & 5 &   446.70 &  -717.13 \\
0.01 & 6 &  -446.77 &  -717.20 \\
0.01 & 7 &   566.15 &  -943.78 \\
0.01 & 8 &  -566.21 &  -943.85 \\
0.01 & 9 &   705.95 & -1110.35 \\
0.01 & 10 & -706.07 & -1110.40 \\
\hline
\end{tabular}
\end{table}

\begin{table}[!h]
\centering
\caption{The first 10 quasinormal modes for $r_+=100$, for various values of $n$.}
\begin{tabular}{|cccc|}
\hline
$n$ & $k$ & Re$[\omega]$ & Im$[\omega]$ \\
\hline
0.1 & 1 &   184.84 &  -266.24 \\
0.1 & 2 &  -185.28 &  -266.59 \\
0.1 & 3 &   316.45 &  -491.83 \\
0.1 & 4 &  -317.02 &  -492.34 \\
0.1 & 5 &   448.28 &  -718.54 \\
0.1 & 6 &  -448.98 &  -719.15 \\
0.1 & 7 &   568.38 &  -946.40 \\
0.1 & 8 &  -569.06 &  -946.99 \\
0.1 & 9 &   719.78 & -1108.06 \\
0.1 & 10 & -720.90 & -1108.30 \\
\hline
\end{tabular}
\qquad
\begin{tabular}{|cccc|}
\hline
$n$ & $k$ & Re$[\omega]$ & Im$[\omega]$ \\
\hline
0.27 & 1 &    -0.09 &  -85.33 \\
0.27 & 2 &    50.58 & -336.78 \\
0.27 & 3 &   -51.54 & -338.07 \\
0.27 & 4 &     1.96 & -560.62 \\
0.27 & 5 &   230.53 & -586.13 \\
0.27 & 6 &  -231.54 & -585.77 \\
0.27 & 7 &   400.83 & -771.17 \\
0.27 & 8 &  -401.35 & -771.17 \\
0.27 & 9 &   545.37 & -945.26 \\
0.27 & 10 & -545.74 & -945.35 \\
\hline
\end{tabular}
\label{Table:overtones4}
\end{table}

%\clearpage
%\subsubsection{Extension to non zero ${\cal N}$}
So far we have assumed ${\cal N}=0$ in  the definition (\ref{C}) of the separation constant $C$ of the angular equation. We can extend our analysis to non zero values of ${\cal N}$ restricting ourselves to ${\cal N} \in [-2,2]$. We study again the behaviour of the lowest QNM for various values of the NUT charge, using $r_+=1$ and $r_+=100$ as indicative horizon values for intermediate and large black holes respectively. The generic behaviour of the QNMs in the ${\cal N} = 0$ case  still hold as it can be seen in table \ref{table:w_non_zero_N},
%figures \ref{fig:w_re_non_zero_N}-\ref{fig:w_norm_non_zero_N}, 
where we notice a similar damping around the same critical values of the NUT charge for all cases. We also notice a slight asymmetry between positive and negative values of ${\cal N}$ for which we do not have an explanation. Clearly this case deserves further investigation which is beyond the scope of the present work. 

\begin{table}[!h]
\centering
\caption{Lowest QNM against different values of the NUT charge for different values of $\cal N$, for indicative horizons.}
\begin{tabular}{|cccccc|}
\hline
$r+$ & $\cal N$ & $n$ & $Re[\omega]$ & $-Im[\omega]$ & $|\omega|$ \\
\hline
1 & -2.0 & 0.10 & 2.82 & 2.25 & 3.61 \\
1 & -2.0 & 0.25 & 0.23 & 1.10 & 1.12 \\
1 & -2.0 & 1 & 0.32 & 3.30 & 3.32 \\
1 & -1.0 & 0.10 & 2.60 & 2.58 & 3.66 \\
1 & -1.0 & 0.25 & 0.11 & 1.44 & 1.44 \\
1 & -1.0 & 1 & 0.11 & 3.38 & 3.38 \\
1 & 1.0  & 0.10 & 2.82 & 2.25 & 3.61 \\
1 & 1.0  & 0.25 & 0.23 & 1.10 & 1.12 \\
1 & 1.0  & 1 & 0.32 & 3.30 & 3.32 \\
1 & 2.0  & 0.10 & 3.41 & 2.01 & 3.96 \\
1 & 2.0  & 0.25 & 0.18 & 0.54 & 0.57 \\
1 & 2.0  & 1 & 0.50 & 3.17 & 3.21 \\
\hline
\end{tabular}
\qquad
\begin{tabular}{|cccccc|}
\hline
$r+$ & $\cal N$ & $n$ & $Re[\omega]$ & $-Im[\omega]$ & $|\omega|$ \\
\hline
100 & -2.0 & 0.10 & 184.42 & 265.88 & 323.58 \\
100 & -2.0 & 0.27 & 0.280 & 85.32 & 85.32 \\
100 & -2.0 & 1.00 & 0.27 & 140.25 & 140.25 \\
100 & -1.0 & 0.10 & 184.84 & 266.24 & 324.11 \\
100 & -1.0 & 0.27 & 0.09 & 85.33 & 85.33 \\
100 & -1.0 & 1.00 & 0.09 & 140.25 & 140.25 \\
100 &  1.0 & 0.10 & 184.42 & 265.88 & 323.58 \\
100 &  1.0 & 0.27 & 0.28 & 85.32 & 85.32 \\
100 &  1.0 & 1.00 & 0.27 & 140.25 & 140.25 \\
100 &  2.0 & 0.10 & 184.01 & 265.51 & 323.04 \\
100 &  2.0 & 0.27 & 0.47 & 85.31 & 85.31 \\
100 &  2.0 & 1.00 & 0.45 & 140.25 & 140.25 \\
\hline
\end{tabular}
\label{table:w_non_zero_N}
\end{table}

%\begin{figure}[!h]%
%    \centering
%    \subfigure[$r_+=1$]{\includegraphics[width=0.45\textwidth]{qnm_real_non_zero_N_intermediate.pdf}}
%    \subfigure[$r_+=100$]{\includegraphics[width=0.45\textwidth]{qnm_real_non_zero_N_large.pdf}}
%    \caption{Real part of the lowest QNM against different values of the NUT charge for different values of $\cal N$, for indicative horizons.}
%    \label{fig:w_re_non_zero_N}
%\end{figure}
%
%\begin{figure}[!h]%
%    \centering
%    \subfigure[$r_+=1$]
%{\includegraphics[width=0.45\textwidth]{qnm_im_non_zero_N_intermediate.pdf}}
%    \subfigure[$r_+=100$]{\includegraphics[width=0.45\textwidth]{qnm_im_non_zero_N_large.pdf}}
%    \caption{Imaginary part of the lowest QNM against different values of the NUT charge for different values of $\cal N$, for indicative horizons.}
%    \label{fig:w_im_non_zero_N}
%\end{figure}
%
%\begin{figure}[!h]%
%    \centering
%    \subfigure[$r_+=1$]
%{\includegraphics[width=0.45\textwidth]{qnm_norm_non_zero_N_intermediate.pdf}}
%    \subfigure[$r_+=100$]{\includegraphics[width=0.45\textwidth]{qnm_norm_non_zero_N_large.pdf}}
%    \caption{Norm of the lowest QNM against different values of the NUT charge for different values of $\cal N$, for indicative horizons.}
%    \label{fig:w_norm_non_zero_N}
%\end{figure}

\clearpage
\section{Discussion}

In this work we presented  numerical results for the quasinormal modes of the massless scalar fluctuations in a Taub-NUT AdS$_4$ background. This work is a natural continuation of \cite{Kalamakis:2020aaj} where we have analytically studied the relevant fluctuation equations. The main issue identified in \cite{Kalamakis:2020aaj} was that the globally regular solutions of the angular equation are not compatible with the presence of complex quasinormal modes. In this work we ignored this issue and considered angular modes that correspond to highest weight $SU(2)$ representations, that are non-unitary and not globally regular. The resulting radial equation does not guarantee the stability of the QNMs. This is reminiscent of the case of odd gravitational fluctuations in Schwarzchild AdS$_4$ \cite{Cardoso:2001bb}. Nevertheless, we are able to show that there exists a region in QNM space that contains only stable modes. The radius of this stable QNM region increases with $r_+$, thus larges BHs are more stable. For fixed $r_+$ the radius of the region shrinks as $n$ increases, but curiously it reaches a constant non-zero value for infinite $n$. 

The main result of our work is the existence of a critical NUT parameter $n_{cr}$ such that for $n>n_{cr}$ the lowest QNMs are overdamped. Remarkably, the value of $n_{cr}$ is rather insensitive to the horizon radius $r_+$ as we were able to argue semi-analytically. Moreover, we observed that the NUT charge induces a fine-splitting of the QNMs which becomes maximum near $n_{cr}$. These observations draw an interesting picture for the holographic fluid of TNAdS$_4$, in particular one in which $n_{cr}$ separates an oscillating from an overdamped phase. We believe that this deserves further study.

An alternative holographic interpretation of Taub-NUT AdS$_4$ can be based on the regular solutions of the angular equation presented in the Appendix A. As we have mentioned, this picture corresponds to the quantization of  $\Omega$, hence there is no ringdown of the TNAdS$_4$ spacetime. We actually find that
\be
\label{monopQ}
2\Omega=4n\omega=k\,,\,\,\,k\in\mathbb{Z^+}\,.
\ee
This is the equivalent of the usual Dirac monopole quantization condition $2ge=k'$, $k'\in\mathbb{Z^+}$ if we identify  $2n\leftrightarrow g$ and $\omega\leftrightarrow e$. From the fluid dynamics point of view this result corresponds to the quantization of the total circulation (\ref{totalfluxTN}) as
\be
\label{qcirc}
C_{total}^{TN}=\left(\frac{2\pi\hbar}{M_{TN}}\right)k\,,\,\,\,M_{TN}=\hbar\omega\,,\,\,\,k\in{\mathbb Z^+}\,.
\ee
This coincides with the well-known  condition for the circulation quantization of superfluid vortices, where $M_{TN}$ is the mass of the condensate, see for example \cite{Fetter:2009zz,QVortices}.  This result prompts us to suggest that the non-singular modes around the TNAdS$_4$ background can be interpreted as quantized vortices.

\section*{Acknowledgements}

A.C.P. would like to acknowledge useful correspondence with M. Casals, R. Davison, B. Goutereaux, J. E. Santos, F. Willenborg, H. Zhang and the warm hospitality of CPHT \`Ecole Polytechnique and Universit\'e de Mons during the completion of this work.

\appendix

\section{Regular solutions of the angular equation}

The analysis of the $SU(2)$ irreps can be alternatively and equivalently performed by studying the angular differential equation (\ref{angularEq2}). For example, for $\Omega=0$ (\ref{angularEq3}) is the associated Legendre equation whose solutions are the  associated Legendre functions. Although the latter can be defined for generic complex values of $\ell$ and $m$ \cite{Legendre}, the quantization of these parameters arises by imposing the condition that the {\it physically acceptable} solutions are the Legendre polynomials which are regular over the whole range of the variable $u\in [0,1]$ or equivalently $\theta\in[0,\pi]$. We will attempt a similar approach to our problem here and set $\Omega\neq 0$ from now on. 
Eq. (\ref{angularEq3})  is a hypergeometric equation with solutions given by 
%\begin{align}
%Y(u)=&\theta(m-\Omega)(1-u)^{\frac{m+\Omega}{2}}u^{\frac{m-\Omega}{2}}C[1] \times \notag \\
%&{}_{2}F_{1}\left(m+\frac{1}{2}(1-\sqrt{1+4\lambda}),m+\frac{1}{2}(1+\sqrt{1+4\lambda}),1+m-%\Omega;u\right) + \notag \\ 
%&\theta(\Omega-m)(1-u)^{\frac{m+\Omega}{2}}u^{-\frac{m-\Omega}{2}}C[2] \times \notag \\
%&{}_{2}F_1\left(\Omega+\frac{1}{2}(1-\sqrt{1+4\lambda}),\Omega+\frac{1}{2}(1+\sqrt{1+4\lambda}),1-m+\Omega;u\right)
%\end{align}
%with $C[1]$ and $C[2]$ some constants.
%Setting $L^2=C=\lambda+\Omega^2 = l(l+1)$, where C stands for the $SU(2)$ Casimir the solution %becomes
\begin{align}\label{SphericalGen}
Y_{\ell {\cal N}\Omega}(u)=&\theta({\cal N})C_1(1-u)^{\frac{{\cal N}+2\Omega}{2}}u^{\frac{\cal N}{2}}{}_{2}F_{1}\left({\cal N}+\Omega-\ell,{\cal N}+\Omega+\ell+1,1+{\cal N};u\right)  \notag \\ 
&+\theta(-{\cal N})C_2(1-u)^{\frac{{\cal N}+2\Omega}{2}}u^{-\frac{\cal N}{2}}{}_{2}F_1\left(\Omega-\ell,\Omega+\ell+1,1-{\cal N};u\right)\,,
\end{align}
where $C_1$, $C_2$ are some generically complex constants. By using the theta functions in (\ref{SphericalGen}) we have ensured the regularity of the solution at $u=0$ ($\theta=0$) where it behaves as $Y(u)\xrightarrow{u\rightarrow 0}u^{|{\cal N}|/2}$. The regularity near $u=1$ ($\theta=\pi$) must be separately considered  and has  physical consequences. We consider  the cases ${\cal N}>0$,  ${\cal N}<0$ and ${\cal N}=0$ separately.  
% \begin{figure}[h]
%\centering
%\includegraphics[scale=0.25]{Singularities.jpeg}
%%\includegraphics[scale=0.7]{USchTNm=0smalln.pdf}\includegraphics[scale=0.7]{USchTNm=0largen.pdf}
%\caption{
%Domains and regularity conditions for $Y_{\ell{\cal N}\Omega}(u)$ when ${\cal N}>0$ and $u\rightarrow 1$.}
%\end{figure}
For ${\cal N}>0$ we have
\begin{align}
\label{YOM}
Y_{\ell {\cal N}\Omega}(u) \xrightarrow{u\rightarrow 1} &(1-u)^{\frac{1}{2}({\cal N}+2\Omega)}\left(-\frac{\pi \csc[\pi({\cal N}+2\Omega)]}{\Gamma(1+{\cal N}+2\Omega)}\frac{\Gamma(1+{\cal N})}{\Gamma(-\ell-\Omega)\Gamma(1+\ell-\Omega)}+ ...\right) + \notag \\
&\hspace{-2cm}+(1-u)^{-\frac{1}{2}({\cal N}+2\Omega)}\left(\frac{(-1)^{-{\cal N}-2\Omega}\pi \csc[\pi({\cal N}+2\Omega)]}{\Gamma(1-{\cal N}-2\Omega)}\frac{\Gamma(1+{\cal N})}{\Gamma(-\ell+{\cal N}+\Omega)\Gamma(1+\ell+{\cal N}+\Omega)}+...\right)\,,
\end{align}
where the dot denote terms proportional to increasing powers of $1-u$. For a given $\ell>0$ we must ensure that there is no singular term in (\ref{YOM}). When ${\cal N}+2\Omega>0$ the second line is singular\footnote{Notice that $\csc(\pi x)/\Gamma(1-x)$ is regular for $x=1,2,3,..$.}  for ${\cal N}+\Omega =m=\ell-k$, $k=0,1,2,3..$. with $\Omega$ and ${\cal N}$ otherwise unrestricted. When ${\cal N}+2\Omega<0$, the first line in (\ref{YOM}) is singular and we must require that $\Omega=-\ell +n$, $n=0,1,2,3...$ with ${\cal N}$ unrestricted. 
%To begin, we consider $m>0$ and then we distinguish two cases: a) when $-m<\Omega<m$ %the second line is singular, and b) when $\Omega<-m$ the first line is singular instead. To get %rid of the singularity in the a) case we must require that $m=\ell,\ell-1,\ell-2,...$. 
%Notice  that $m$ can be integer (in which case 0 is included) or half-integer (in which case 0 is not incuded). 
%In case b), we must instead require $\Omega=-\ell,-\ell+1,-\ell+2,...$, with $-\ell\leq\Omega <-%m$. 
 %When $m<0$ only the first term in (\ref{YOM}) is singular and to get rid of it we have to require %that $\Omega=-\ell,-\ell+1,-\ell+2,...$. The result is that the regularity $Y_{\ell m\Omega}(\theta,%\phi)$ over the whole parameter region $\theta\in[0,\pi]$ %(or equivalently, the requirement that %$Y_{\ell m\Omega}(\theta,\phi)$ are {\it sections} over line bundles) 
%connects $\Omega=2n\omega$ to $\ell$ and $m$.
On top of regularity, we want to require the single-valuedness of the eigenfunctions $Y_{\ell m\Omega}(\theta,\phi)$ under the rotation $\phi\rightarrow \phi +2\pi n $, $n\in\mathbb{Z}$. This gives $m-\Omega={\cal N}\in \mathbb{Z}$ which combined with the previous results yields  $m-\Omega =2\ell -k+n={\cal N}\in {\mathbb Z}$. For generic integers $k,n$ this implies that  $\ell=M/2$, $M\in {\mathbb Z}$, namely we find that  the $SU(2)$ eigenvalues are half-integers or integers when the corresponding eigenfunctions are regular and single-valued. Of course, when $\Omega=0$ we return to the textbook result that requires $\ell$ to be integer for regular and single-valued eigenfunctions of the Laplacian on the two-sphere. 

%Next, if we further require that the eigenvalues of the $SU(2)$ Casimir equation (\ref{angularEq1}) are a normalizable set, or equivalently they form a finite-dimensional irrep of $SU(2)$, then it is well-known that $\ell=2k$ with $k\in \mathbb{Z}^+$, namely a positive integer of  half-integer. Then, requiring the the eigenvalues are regular everywhere the analysis above gives that $m$ and $\Omega$ are integers or half-integers with $-\ell\leq \Omega<m\leq \ell$. Depending on whether $\ell>0$ is an integer of a half-integer, $m$ and $\Omega$ span correspondingly all integer or half-integers in the interval $[-\ell,\ell]$. 

For  $\Omega > m$ the solution near $u=1$ behaves as 
\begin{align}
\label{YMO}
Y_{\ell {\cal N}\Omega}(u) \xrightarrow{u\rightarrow 1} &(1-u)^{\frac{1}{2}(m+\Omega)}\left(-\frac{\pi \csc[\pi(m+\Omega)]}{\Gamma(1+m+\Omega)}\frac{\Gamma(1-m+\Omega)}{\Gamma(-m-\ell)\Gamma(1-m+\ell)}+ ...\right) + \notag \\
&(1-u)^{-\frac{1}{2}(m+\Omega)}\left(\frac{\pi \csc[(-1)^{-m-\Omega}\pi(m+\Omega)]}{\Gamma(1-m-\Omega)}\frac{\Gamma(1-m+\Omega)}{\Gamma(-\ell+\Omega)\Gamma(1+\ell+\Omega)}+...\right)\,.
\end{align}
This is obtained from (\ref{YOM}) under the map $m\leftrightarrow \Omega$ which implies that our analysis will yield $m$ and $\Omega$ to be integers {\it or} half-integers satisfying $-\ell\leq m<\Omega\leq \ell$, depending on whether $\ell>0$ is correspondingly an integer or a half-integer. 

 For $\Omega=m$ there is no $\phi$ dependence in (\ref{PhiSep}), nevertheless there is still a non-trivial "magnetic" quantum number since  the solution becomes
\be
\label{YO=M}
Y_{\ell {\cal N}\Omega}(u)\xrightarrow{{\cal N}\rightarrow 0}=C(1-u)^{m}{}_2F_1(-\ell+m,1+\ell+m;1,u)\,.
\ee
Requiring this to be regular at $u=0,1$ we obtain again that $m$ is an integer {\it or} a half-integer with $|m|\leq \ell$.

To summarize, we have found regular solutions of the angular equation (\ref{angularEq2}) which are single valued as $\phi\mapsto\phi+2n\pi$, $n\in\mathbb{Z}$, under the requirement that $m$ and $\Omega$ are integers {\it or} half-integers satisfying $-\ell\leq m,\Omega\leq\ell$, with $\ell=0,1/2,1,3/2,2,..$. Explicitly, the general solution of the scalar fluctuation equation (\ref{Phieom}) is given by
\be
\label{GenSolPhi}
\Phi(t,r,\theta,\phi)=e^{-i\omega t}e^{i{\cal N}\phi} Y_{\ell {\cal N}\Omega}(\theta)R(r)\,,\,\,\,\,{\cal N}=m-\Omega\in \mathbb{Z}\,.
\ee
%This is regular as $\theta\rightarrow 0,\pi$ and periodic as $\phi\rightarrow\phi+2\pi$. To achieve that we had to impose the quantization condition (\ref{monopQ}). 
Notice now that due to (\ref{monopQ}) the solution (\ref{GenSolPhi}) is periodic in the time coordinate under shifts of the form
\be
\label{pertime}
t\rightarrow t+8\pi nk\,,\,\,\,\,\,k\in\mathbb{Z}\,.
\ee
This is exactly the usual Misner-string periodicity condition \cite{Misner:1963fr,Astefanesei:2004kn}, and we are describing a smooth $U(1)$ bundle. 

\bibliographystyle{ieeetr}

\bibliography{Refs}

%\section{Introduction}
\end{document}